\documentclass[letterpaper]{article}

\usepackage[T1]{fontenc}

\usepackage{geometry}
\usepackage{setspace}

\usepackage[articletitle=true,doi=true]{achemso}

\usepackage{graphicx}
\usepackage{float}
\newfloat{scheme}{htbp}{los}
\floatname{scheme}{Scheme}
\floatname{chart}{Chart}
\newfloat{graph}{htbp}{loh}

\usepackage{chemformula} 
\usepackage[version = 4]{mhchem} 

\usepackage{orcidlink}    
\usepackage{fontawesome5} 
\newcommand{\orcidicon}[1]{\href{https://orcid.org/#1}{\textsuperscript{\faOrcid}}}

\usepackage{amsmath,amssymb,amsfonts}

\usepackage{xurl,url}
\usepackage{hyperref}
\hypersetup{
    colorlinks = true,
    breaklinks = true,
    linkcolor = cyan,
    urlcolor = cyan,
    citecolor = cyan,
    filecolor = cyan,
    allcolors = cyan
}
\usepackage[nameinlink]{cleveref}
\crefname{section}{Section}{Sections}
\Crefname{section}{Section}{Sections}
\crefname{figure}{Figure}{Figures}
\Crefname{figure}{Figure}{Figures}
\crefname{table}{Table}{Tables}
\Crefname{table}{Table}{Tables}
\crefname{equation}{Eq.}{Eqs.}
\Crefname{equation}{Eq.}{Eqs.}

\newcounter{mainappendix}
\crefname{mainappendix}{Appendix}{Appendices}
\Crefname{mainappendix}{Appendix}{Appendices}

\crefformat{mainappendix}{#2Appendix#3}
\Crefformat{mainappendix}{#2Appendix#3}

\usepackage[ruled]{algorithm2e}
\SetKwComment{Comment}{/* }{ */}
\SetKwFor{While}{while}{do}{end{\hspace{0.25em}}while}
\SetKwInput{KwRequire}{Prior Information}
\SetKwInput{KwParam}{Configuration Parameters}

\crefname{algocf}{Algorithm}{Algorithms}
\Crefname{algocf}{Algorithm}{Algorithms}

\usepackage{booktabs,threeparttable}
\usepackage{caption,subcaption}

\usepackage{authblk}
\author[1]{Deepanjhan Das\textsuperscript{\orcidlink{0009-0004-4295-8396}}}
\author[1]{Shankar Narasimhan*\textsuperscript{\orcidlink{0000-0002-0558-0206}}}
\affil[1]{Department of Chemical Engineering, Indian Institute of Technology Madras, Chennai-600036, India}

\title{A recursive subspace based method for errors-in-variables model identification of time-varying systems}
\date{*Email: naras@iitm.ac.in}

\begin{document}

\maketitle

\begin{abstract}
  The Subspace-based Model Identification algorithm using a modified Iterative Principal Component Analysis (SMI-IPCA) is a theoretically rigorous method for identifying a linear state-space model of a multi-input multi-output (MIMO) process, in an errors-in-variables (EIV) setting.  The method can simultaneously estimate unknown heteroskedastic noise variances corrupting the input and output measurements, along with the state space model. This work proposes a recursive formulation of SMI-IPCA (RSMI-IPCA) enabling online identification and adaptive model updates as and when new data arrive. By maintaining a fixed length lag window rather than storing the complete historical data, RSMI-IPCA estimates measurement noise variances, process order, while simultaneously identifying the state-space matrices, making it suitable to monitor time-varying systems, whether the induced changes are slow or abrupt. The algorithm gradually adapts to slow sensor degradation (time-varying noise variances), changes in process operating conditions (time-varying model parameters), and structural modifications (varying model order). Simulation studies are presented to demonstrate the efficacy and practical applicability of the proposed algorithm.
\end{abstract}

\section*{Keywords}
  Time varying MIMO systems, Recursive subspace identification, Adaptive system identification, Errors-in-variables identification



\section{Introduction} \label{sec:intro}
The identification of linear state-space models from experimental or operational data is fundamental to modern control engineering and has been the subject of extensive research over several decades. Such models are essential for developing feedback controllers, state estimators, and predictive algorithms across diverse applications ranging from chemical processes~\cite{Garcia:1989} and manufacturing systems to aerospace vehicles~\cite{Kim:2023} and power systems~\cite{Monticelli:1999}. One of the classical system identification methods includes prediction error method (PEM) \cite{Stoica:1989,Ljung:1999,Tangirala:2015}, that has been extensively developed and refined. However, subspace-based model identification (SMI) approaches have emerged as powerful alternatives to PEM, offering superior numerical robustness and computational efficiency particularly for multivariable systems. Foundational methods such as the multivariable output error state-space (MOESP) \cite{Verhaegen:1992a,Verhaegen:1992b}, and numerical algorithms for subspace-based state-space system identification (N4SID) \cite{Overschee:1994} map the dynamic identification problem to an equivalent static subspace problem through a clever use of data stacking and orthogonal projections, enabling the use of well-conditioned linear algebra operations rather than nonlinear optimization. However, these methods are built upon a restrictive and often unrealistic assumption that the measured input variables are free from errors, with noise affecting only the measured output variables. This assumption is reasonable when experiments are carefully designed and inputs are generated by precise actuators under complete experimental control. Nonetheless, in the vast majority of practical scenarios, particularly when identification is performed on continuously operating processes, when data are obtained from industrial sensor measurements, or when the system is subjected to environmental disturbances, both input and output measurements inevitably contain errors. This situation is referred to as the errors-in-variables (EIV) problem and represents a significant departure from the classical identification framework. 

The presence of measurement errors in both inputs and outputs introduces three interconnected challenges that are not adequately addressed by classical identification methods. First, the standard least-squares approach and many classical subspace-based methods yield biased parameter estimates in presence of noisy input measurements. The bias arises because errors in the measured input correlate with the regression residuals, violating the independence assumption implicit in classical methods, as pointed out by S{\"o}derstr{\"o}m~\cite{Soderstrom:1981}. Beyond the bias problem, accurate estimation of measurement noise variances becomes essential for practical applications. Knowledge of these variances is fundamental to performing statistical significance tests on estimated parameters \cite{Stoica:1989,Ljung:1999}, designing robust state estimators such as Kalman filters that account for actual measurement quality \cite{Kalman:1960,Grewal:2015}, and implementing uncertainty-aware control algorithms \cite{Astrom:1970,Stengel:1994}. Among several proposed methods (well documented in \cite{Soderstrom:2018}) to address the identification of EIV dynamical models, the instrumental variable (IV) approach is arguably the most widely used \cite{Stoica:1995,Chou:1997,Li:2001,Wang:2002}, which neutralizes the effect of noise in input and output measurements. In contrast to IV-based approaches, that eliminate the requirement of estimating the noise variances, a different class of proposed methods include the bias elimination least squares (BELS) \cite{Zheng:2002}, bias compensated least squares (BCLS) \cite{Ikenoue:2005}, which estimate the noise variances to rectify the parameter bias resulting from the use of ordinary least squares method for EIV problem. Later, S{\"o}derstr{\"o}m \cite{Soderstrom:2012} generalized the IV method by combining it with BCLS for EIV MIMO systems. 

However, these methods require the input and output process orders to be specified, or they have to be estimated using criteria such as Akaike's Information Criterion (AIC)~\cite{Akaike:1998} and Bayesian Information Criterion (BIC)\cite{Schwarz:1978}, which leads to the second challenge. Classical order selection criteria (AIC or BIC), derived under the assumption of error-free inputs, become unreliable when input measurements are corrupted by noise and often lead to significant overestimation or underestimation of the true process order. Although there exists adapted versions of these criteria under restrictive assumptions \cite{Li:2001}, the absence of a theoretically rigorous framework for joint determination of both process order and noise variance estimation compounds this difficulty substantially. Thirdly, in industrial processes and real-world systems, dynamics and measurement noise characteristics change over time due to sensor degradation through corrosion, bio-fouling, scaling, or electronic drift; changes in operating conditions; structural modifications to the process such as opening or closing bypass valves, bringing standby equipment online, and aging of equipment. Identifying such time-varying systems requires not only an identification method capable of handling the EIV setting, but one that can adapt recursively in real-time without the need to store and reprocess complete historical data. Tracking variance evolution in real-time enables predictive maintenance scheduling and helps distinguish between genuine process changes and measurement quality deterioration \cite{Hashlamon:2016,Hajiyev:2025}, allowing process operators to plan maintenance proactively rather than reactively. For systems with limited resources, such as embedded systems, edge devices, wireless sensor networks, and Internet-of-Things applications, batch identification methods are impractical due to constraints on data storage and computational capacity \cite{Chuang:2019,Venkatesh:2025,Fuente:2025}. In such environments, the recursive nature of the identification algorithm is essential, not merely convenient.

In past few decades, numerous recursive subspace identification (RSID) methods have been proposed for various problem classes. Notably, Lovera et al.~\cite{Lovera:2000} proposed recursive extensions of MOESP including a recursive past output EIV (PO-EIV) variant using instrumental variables. While this pioneering work demonstrated that EIV identification could be performed recursively and adaptively, it requires the order to be known \textit{a priori} and does not address the variance estimation. Jiang and Fang~\cite{Jiang:2009} developed a RSID for closed-loop stochastic systems, though this method operates in the classical regime without variance estimation capability and requires the system order to be pre-specified. Weng and Loh~\cite{Weng:2011} proposed a RSID with forgetting factors for structural modal tracking, enabling faster convergence during system changes, but again remaining in the classical domain, and several user-defined parameters including the model order to be specified. De Cock et al.~\cite{Cock:2006} presented RSID algorithms for in-flight modal analysis of aircraft, demonstrating practical applicability to real-world systems and accommodating both process and measurement noise, though operating in the output error domain without explicit input measurement noise treatment and with process order assumed known or overestimated. Alenany and Shang~\cite{Alenany:2013} proposed a RSID with incorporation of exact prior information via constrained least squares (CLS), estimating the system order heuristically by evaluating the number of significant singular values. Oku et al.~\cite{Oku:2001} developed a change detection scheme within the past input MOESP (PI-MOESP) recursive framework to track temporal variations in system dynamics, with an important capability being the estimation of output measurement variances, though confined to the classical domain and requiring the system order to be known. Houtzager et al.~\cite{Houtzager:2012}, representing the state-of-the-art (SOTA) in RSID, proposed recursive predictor-based subspace identification (RPBSID) approaches based on the optimized version of PBSID ($\text{RPBSID}_{\text{opt}}$), and propagator method ($\text{RPBSID}_{\text{pm}}$) with sophisticated applications to real-time closed-loop tracking, yet this approach also operates within the classical non-EIV framework. Hou et al.~\cite{Hou:2017} addressed recursive identification of systems subject to time-varying load disturbances, an important practical problem, though the focus is on handling specific disturbance characteristics within the output-error regime. Bathelt et al.~\cite{Bathelt:2017} developed a theoretically rigorous coordinate-free framework for RSID with convergence guarantees and demonstrated application in EIV setting, though the method does not address the critical problems of process order determination or measurement noise variance estimation. A comprehensive survey of this literature reveals a consistent pattern: while these methods collectively advance the SOTA in RSID, they either operate in the classical domain, or if they do handle EIV, they do not simultaneously estimate noise variances and determine process order through principled statistical methods. The absence of variance estimation is particularly problematic because it prevents deployment of these methods in applications requiring sensor quality assessment, fault diagnosis, and robust state estimation.

A significant advancement in EIV state-space identification was achieved through the subspace-based model identification using iterative principal component analysis (SMI-IPCA), proposed by Ramnath and Narasimhan~\cite{Keerthan:2023}, representing the first systematic approach to simultaneously address the three challenges of noise variance estimation, theoretical order determination, and EIV treatment. The theoretical rigor and comprehensive treatment of the EIV problem in SMI-IPCA represent a significant advancement in system identification methodology. However, SMI-IPCA is fundamentally a batch method requiring access to all historical data to perform its identification, necessitating complete data storage and reprocessing whenever updates are required. This creates a critical bottleneck for time-varying systems and renders the method impractical for many real-time applications. More recently, Pradeep and Narasimhan~\cite{Pradeep:2025} proposed recursive IPCA (RIPCA), extending the batch IPCA algorithm to recursive operation. By reformulating the optimization objective for variance estimation to depend only on the covariance matrix rather than the complete data matrix, RIPCA achieves true online model and variance estimation with memory footprint dependent only on the number of variables and independent of data history length. RIPCA successfully demonstrates recursive variance tracking that follows time-varying measurement noise variances without storing past data, online constraint model updates that adapt the linear relationships as operating conditions change, and convergence to batch estimates under constant system conditions. However, RIPCA is restricted to static constraint identification models where the underlying formulation addresses linear equations relating variables.

The review of the literature on identification of linear dynamic EIV MIMO processes reveals that a recursive method for identifying a time varying process and/or time varying noise variances and system order is not currently available. This work proposes such an approach (RSMI-IPCA) by extending the SMI-IPCA method from batch to fully recursive implementation. By carefully managing the interaction between dynamic constraint identification and variance estimation through iterative procedures that operate on recursively updated covariance matrices rather than historical data, RSMI-IPCA achieves online adaptation to time-varying systems without sacrificing the theoretical rigor of the batch method.

The remainder of this paper is organized as follows. \Cref{sec:background} provides the theoretical foundations for RPCA, RIPCA, including the formulation of the EIV state-space identification problem and presentation of the batch SMI-IPCA algorithm. \Cref{sec:rsmi-ipca} develops the proposed recursive variant, detailing the key modifications required to enable recursive operation while maintaining theoretical rigor and demonstrating how the iterative optimization for variance estimation can be reformulated to depend only on covariance matrices. \Cref{sec:simulation} demonstrates the effectiveness of RSMI-IPCA through simulation studies on benchmark systems, illustrating its capability to track time-varying noise and system structure. Finally, \Cref{sec:conclusion} provides concluding remarks and discusses directions for future research.

\section{Background} \label{sec:background}
We provide a brief discussion on Recursive PCA (RPCA), that broadens the model identification applicability of PCA for the EIV case, by adapting to the changes in the process. This is followed by a brief review of the Recursive Iterative PCA algorithm (RIPCA), which extends PCA, by combining RPCA and the Iterative PCA (IPCA) framework to handle heteroskedastic noise, and to monitor the time-varying steady-state process. Subsequently, we critically review the SMI-IPCA algorithm, that is used to identify state space models of dynamic LTI MIMO processes. 

\subsection{Recursive PCA} \label{subsec:rpca}
Principal Component Analysis (PCA) is a statistical technique that is primarily used for data compression \cite{Abdi:2010,Dai:2012}. However, it can also be used to identify linear relations that relate the variables, known as model identification \cite{Jolliffe:2002,ShankarSir:2008} in the EIV case, when the errors have identical variances (homoskedastic case). The identification problem asks for the estimation of the constraint matrix $\mathbf{A} \in \mathbb{R}^{d\times n}$, which relates $n$-dimensional \textit{noise-free} variables $\mathbf{z}_n^*(k)$ at instant $k$ by $d$, ($<n$) linearly independent constraints, given the $N$ samples of corresponding noisy measurements $\mathbf{z}_n(k)$. This can be compactly written as:
\begin{subequations}
    \label{eq:2.1}
    \begin{align}
        &\mathbf{A}(\mathbf{Z}_n^*)^\intercal = \mathbf{0}_{d\times N}^{} \label{eq:2.1a} \\
        &\mathrm{subject \ to \ the \ measurement \ noise} \nonumber \\
        &\mathbf{Z}_n^{} = \mathbf{Z}_n^* + \mathbf{E}_n^{} \label{eq:2.1b}    
    \end{align}
\end{subequations}
where, $\mathbf{Z}_n$ is the data matrix, defined as $\mathbf{Z}_n = [\mathbf{z}_n(1) \ \mathbf{z}_n(2) \ \ldots$ $\mathbf{z}_n(N)]^\intercal \ \in \mathbb{R}^{N\times n}$. The true values $\mathbf{z}_n^*(k)$ corresponding to each sample are assumed to be deterministic and bounded by their first and second moments \cite{ShankarSir:2008}. The error matrix $\mathbf{E}_n$ is also constructed in a similar way using the error vectors $\mathbf{e}_n(k)$, which are assumed to be Gaussian white noise with zero mean and a scale of the identify matrix, that is, $\sigma_e^2\mathbf{I}_n$, corrupting the measurements across all the variables and samples. Furthermore, the errors in different variables and samples are assumed to be mutually independent and also independent of the true values of the variables.

\Cref{eq:2.1a} implies that $\mathbf{A}$ spans the null space of $(\mathbf{Z}_n^*)^\intercal$ which has a rank $n-d$. However, having access only to the noisy data matrix $\mathbf{Z}_n$, a basis for the null space is determined from the eigenvectors corresponding to eigenvalues having magnitude equal to $\sigma_e^2$ of the sample covariance matrix $\mathbf{S_Z}_n$ of $\mathbf{Z}_n$ around zero. $\mathbf{S_Z}_n$ is an unbiased estimate of the covariance matrix of the noisy measurements $\mathbf{Z}_n$ which satisfies the following relation:
\begin{align}
    \mathbf{\Sigma_Z}_n &= \mathbb{E}\left(\mathbf{S_Z}_n\right) = \mathbb{E}\left(\frac{1}{N} \mathbf{Z}_n^\intercal \mathbf{Z}_n \right) \nonumber \\
    & = \mathbf{\Sigma}_{\mathbf{Z}_n^*} + \sigma_e^2 \mathbf{I}_n \label{eq:2.2}
\end{align}
where, $\mathbb{E}(\cdot)$ is the mathematical expectation operator. Henceforth, in this work we refer to $\mathbf{S_Z}_n$ of noisy measurements around zero simply as the covariance matrix. 

Recursive PCA, proposed by Li et al.~\cite{Li:2000}, on the other hand has been developed for adapting to the changes in the steady-state process by updating the model recursively while receiving new measurements. The adaptation implies the recursive update in the model coefficients, which can be achieved using two approaches. The naive approach updates the mean and covariance matrix of the past data recursively using the complete past data and newly obtained data, whereas the comparatively more optimized approach requires only the updated covariance matrix. Let $\mathbf{Z}_{n,k-1} \in \mathbb{R}^{N_{k-1}\times n}$ be the past data matrix at time $k$, and $\mathbf{S_Z}_{n,k-1}$ be the corresponding covariance matrix. Now, with the newly received data matrix $\mathbf{Z}_{n,k} \in \mathbb{R}^{n_k \times n}$ at time $k$, Li et al.~\cite{Li:2000} give the following general form to update the covariance matrix which weighs the old and new data differently using a forgetting factor $\mu$.
\begin{equation}
    \label{eq:2.3}
    \mathbf{S_Z}_{n,k} = \mu \mathbf{S_Z}_{n,k-1} + \frac{1-\mu}{n_k} \left( \mathbf{Z}_{n,k}^\intercal \mathbf{Z}_{n,k} \right)
\end{equation}
where, $0 < \mu \leq N_{k-1}/N_k < 1$ and $N_k = N_{k-1} + n_k$. If $\mu = N_{k-1}/N_k$, \Cref{eq:2.3} reduces to a simplified update equation which is as follows:
\begin{equation}
    \label{eq:2.4}
    \mathbf{S_Z}_{n,k} = \frac{N_{k-1}}{N_k} \mathbf{S_Z}_{n,k-1} + \frac{1}{N_k} \left( \mathbf{Z}_{n,k}^\intercal \mathbf{Z}_{n,k} \right)
\end{equation}
here, it may be noted that if RPCA is applied at each sampling instant $k$, $\mathbf{Z}_{n,k}$ will correspond to the measurement vector received at the time $k$ and $n_k$ will be unity. 

The eigenvalues and eigenvectors of the updated covariance matrix $\mathbf{S_Z}_{n,k}$ can be obtained by performing spectral decomposition, from which the number of constraints and the linear constraint model can be estimated. The use of spectral decomposition of the covariance matrix $\mathbf{S_Z}_{n,k}$ is a major advantage over the use of SVD of $\mathbf{Z}_n$, which leads to a significant reduction in the storage requirements. Further enhancement of the methodological efficiency states that the eigenvalues and eigenvectors themselves can be directly updated, despite updating the covariance matrix, which involves two rank-one modifications of a symmetric matrix, as noted from \Cref{eq:2.3}. As opposed to finding the spectral decomposition of a $n\times n$ square matrix, which involves $\mathcal{O}\left(n^3\right)$ operations, the rank-one update of the eigenvalues and eigenvectors of symmetric matrices \cite{Golub:1983,Bunch:1978} require only $\mathcal{O}\left(n^2\right)$ operations. 

\subsection{Recursive IPCA} \label{subsec:ripca}
PCA assumes homoskedasticity and uncorrelated errors in order to solve the identification problem, whereas the general scenario comprises different errors across both variables and samples, and could also be correlated. Wentzell et al.~\cite{Wentzell:1997} proposed a Maximum Likelihood PCA (MLPCA) where, given the complete knowledge of the noise variances and covariances, the model can be obtained under very general noise characteristics. A simpler yet practically useful consideration leads to mutually independent errors which are different in different variables, but same across the samples, and independent of the true values of variables. Therefore, \Cref{eq:2.2} can be replaced by noise covariance matrix $\mathbf{\Sigma}_e$ as:
\begin{align}
    \mathbf{e}(k) \sim \mathcal{N}(\mathbf{0}, \mathbf{\Sigma}_e) \implies \mathbf{\Sigma_Z}_n = \mathbf{\Sigma}_{\mathbf{Z}_n^*} + \mathbf{\Sigma}_e \label{eq:2.5}
\end{align}

An Iterative PCA (IPCA) algorithm was proposed by Narasimhan and Shah~\cite{ShankarSir:2008}, which simultaneously estimates the unknown $\mathbf{\Sigma}_e$ and identify the model by iterating between the following two major steps: (i) estimation of the constraint matrix $\mathbf{A}$ up to a rotation from a given estimate of $\mathbf{\Sigma}_e$, and (ii) estimation of $\mathbf{\Sigma}_e$ from the given estimate of $\mathbf{A}$. The key idea is to scale the data matrix $\mathbf{Z}_n$ with $\mathbf{\Sigma}_e^{-1/2}$, yielding $\mathbf{Z_S}_n \triangleq \mathbf{Z}_n\mathbf{\Sigma}_e^{-1/2}$. Therefore, \Cref{eq:2.5} takes the following form for the scaled data matrix $\mathbf{Z_S}_n$:
\begin{equation}
    \label{eq:2.6}
    \mathbf{\Sigma_{Z_S}}_n = \mathbb{E}\left( \mathbf{S_{Z_S}}_n \right) = \mathbf{\Sigma}_{\mathbf{Z}^*_{\mathbf{S}_n}} + \mathbf{I}_n
\end{equation}
implying that the smallest $d$ eigenvalues of $\mathbf{\Sigma_{Z_S}}_n$ are unity. Hence, an estimate of the rows of $\mathbf{A}$ can be obtained from the eigenvectors corresponding to the $d$ unity eigenvalues. 

IPCA is applicable for identifiable systems, which requires the following condition to be satisfied:
\begin{equation}
    \label{eq:2.7}
    d(d+1) \geq 2M
\end{equation}
where, $M$ is the number of elements that need to be estimated in $\mathbf{\Sigma}_e$ and for diagonal covariance matrix $M = n$, the number of variables in the system. Given this identifiability condition satisfied, $\mathbf{\Sigma}_e$ can be estimated by maximizing the likelihood of the constraint residuals $\mathbf{r}(k)$, defined as $\mathbf{r}(k) \triangleq \mathbf{\hat{A}z}_n(k)$, which can be shown to have normal distribution with zero mean and covariance $\mathbf{\Sigma}_r = \mathbf{\hat{A}\Sigma}_e \mathbf{\hat{A}}^\intercal$, if $\mathbf{\hat{A}}$ is row equivalent to $\mathbf{A}$. Here, $\mathbf{\hat{A}}$ is a known estimate of the constraint matrix $\mathbf{A}$. Maximizing the log-likelihood of the joint density function of $\mathbf{r}(k)$ leads to the following non-linear optimization problem for estimating $\mathbf{\Sigma}_e$:
\begin{equation}
    \label{eq:2.8}
    \mathbf{\hat{\Sigma}}_e = \underset{\mathbf{\Sigma}_e}{\mathrm{arg\ min}} \left[ N\ \mathrm{log}\left| \mathbf{\Sigma}_r \right| + \sum_{k=1}^N \mathbf{r}(k)^\intercal \left( \mathbf{\Sigma}_r \right)^{-1} \mathbf{r}(k) \right]
\end{equation}

The convergence of the IPCA algorithm is determined by the relative change in the sum of smallest $d$ eigenvalues, where the number of constraints $d$ is determined by successively testing decreasing values from a maximum possible guess value $n-1$ until the smallest $d$ eigenvalues of $\mathbf{S_{Z_S}}_n$ are equal after convergence. Thus, it is easy to validate $d$ using a hypothesis test for testing equality of eigenvalues \cite{Keerthan:2023}.

To address the required adaptations in time-varying steady-state linear models relating the variables, due to changes in operating conditions and to monitor the changes in noise variances because of gradual sensor degradation, Pradeep and Narasimhan~\cite{Pradeep:2025} proposed a Recursive IPCA (RIPCA) approach, which does not require the past data to be stored. This follows from the steps of IPCA, where, for estimating the constraint model, the eigenvalues and eigenvectors are required, for which we make use of the covariance matrix of the scaled data. Given an estimate of the noise covariance $\mathbf{\hat{\Sigma}}_{e,k-1}$ at time instant $k$, the covariance matrix of scaled measurements can be computed as:
\begin{align}
    \mathbf{S_{Z_S}}_{n,k} = \mathbf{\hat{L}}_{\mathbf{e},k}^{-\intercal} \left(\mathbf{S_{Z}}_{n,k}\right) \mathbf{\hat{L}}_{\mathbf{e},k}^{-1} \label{eq:2.9}
\end{align}
where, $\mathbf{\hat{L}}_{\mathbf{e},k}$ is the Cholesky factor of $\mathbf{\hat{\Sigma}}_{e,k-1}$, defined by $\mathbf{\hat{\Sigma}}_{e,k-1} = \mathbf{\hat{L}}_{\mathbf{e},k} \mathbf{\hat{L}}_{\mathbf{e},k}^\intercal$. This is followed by the estimation of noise covariance matrix by solving the optimization problem in \Cref{eq:2.8}, given an estimate of $\mathbf{\hat{A}}_k$ at time instant $k$, which although appears to require the entire data to compute the objective function, can be recast in a form that requires only the scaled, sample covariance matrix in place of the data matrix. The trace of the objective function is used to achieve this, which makes it a scalar. From cyclic property of trace, the second term on the RHS of \Cref{eq:2.8} can be rewritten as:
\begin{align}
    \mathrm{Tr}\left( \sum_{k=1}^{N_k} \left( \mathbf{r}(k)^\intercal \left( \mathbf{\Sigma}_r \right)^{-1} \mathbf{r}(k) \right) \right) &= \mathrm{Tr}\left( \sum_{k=1}^{N_k} \left( \left( \mathbf{\Sigma}_r \right)^{-1} \mathbf{r}(k) \mathbf{r}(k)^\intercal \right) \right) \nonumber \\
    &= \mathrm{Tr}\left( \sum_{k=1}^{N_k} \left( \left( \mathbf{\hat{A}}_k \mathbf{\hat{\Sigma}}_{e,k-1} \mathbf{\hat{A}}_k^\intercal \right)^{-1} \mathbf{r}(k) \mathbf{r}(k)^\intercal \right) \right) \label{eq:2.10}
\end{align}

The first term of \Cref{eq:2.8} is already a scalar. Now, using the definition of the constraint residuals $\mathbf{r}(k)$, the optimization problem can be rewritten as:
\begin{equation}
    \label{eq:2.11}
    \mathbf{\hat{\Sigma}}_{e,k} = \underset{\mathbf{\hat{\Sigma}}_{e,k-1}}{\mathrm{arg\ min}} \left[ N_k\ \mathrm{log}\left| \mathbf{\hat{A}}_k \mathbf{\hat{\Sigma}}_{e,k-1} \mathbf{\hat{A}}_k^\intercal \right|\ + N_k\ \mathrm{Tr} \left( \left( \mathbf{\hat{A}}_k \mathbf{\hat{\Sigma}}_{e,k-1} \mathbf{\hat{A}}_k^\intercal \right)^{-1} \mathbf{\hat{A}}_k \mathbf{S_Z}_{n,k} \mathbf{\hat{A}}_k^\intercal \right) \right] 
\end{equation}
which only requires the updated sample covariance matrix and not the past data. The updated estimate of the noise covariance matrix can now be used to update the sample covariance matrix and iteration between these two steps is essentially the main idea of RIPCA algorithm. This is to note that the unavailability of the noise covariance matrix, which can change with time, restricts RIPCA from using direct rank-one update of the eigenvectors of the transformed sample covariance matrix $\mathbf{S_{Z_S}}_n$, unlike in RPCA.

\subsection{SMI-IPCA essentials} \label{subsec:smiipca}
Prior to the discussion of the proposed recursive adaptation for monitoring dynamic systems, we discuss the essentials underpinning the identification procedure of linear state space model \cite{Keerthan:2023} of a dynamic process in the EIV setting, which has following general mathematical representation:
\begin{subequations}
    \label{eq:2.12}
    \begin{align}
        &\mathbf{x}(k+1) = \mathbf{Ax}(k) + \mathbf{Bu}^*(k) + \mathbf{p}(k) \\
        &\mathbf{y}^*(k) = \mathbf{Cx}(k) + \mathbf{Du}^*(k) 
    \end{align}
\end{subequations}
subject to the measurement noise
\begin{equation}
    \label{eq:2.13}
    \mathbf{y}(k) = \mathbf{y}^*(k) + \mathbf{v}(k); \quad \mathbf{u}(k) = \mathbf{u}^*(k) + \mathbf{w}(k)
\end{equation}
where, $\mathbf{x}(k) \in \mathbb{R}^\eta$ are the state variables, $\mathbf{u}(k) \in \mathbb{R}^\ell,$ and $\mathbf{y}(k) \in \mathbb{R}^m$ are the measurements of inputs and output variables, respectively, whereas the superscript $(\cdot)^*$ denotes the true values of these variables. $\mathbf{p}(k), \mathbf{v}(k),$ and $\mathbf{w}(k)$ are process, output, and input measurement noises, respectively, which are assumed to be mutually independent Gaussian white noise sequences, with zero mean and noise covariances $\mathbf{\Sigma}_p, \mathbf{\Sigma}_v,$ and $\mathbf{\Sigma}_w$, respectively. However, the inclusion of process noise leads to colored noise affecting the output measurements. For simplicity, the noise covariance matrices are further assumed to be diagonal. Some additional, yet important assumptions regarding the system are discussed in \cite{Wang:2002, Keerthan:2023}. 

In Subspace-based Model Identification (SMI) of the dynamic state space model, the problem is mapped to constraint model identification of a static process, by means of stacking lagged measurements of outputs and inputs. Henceforth, the lagged output measurements is denoted by $\mathbf{y}_f(k)$ for a stacking length $f$, which is constructed as:
\begin{equation}
    \label{eq:2.14}
    \mathbf{y}_f(k) = 
        \begin{bmatrix}
            \mathbf{y}(k) \\ \mathbf{y}(k+1) \\ \vdots \\ \mathbf{y}(k+f-1) 
        \end{bmatrix} \in \mathbb{R}^{mf}
\end{equation}
while similar stacking procedure is followed for input measurements, and noise at $k$'th instant denoted by $\mathbf{u}_f(k),$ $\mathbf{p}_f(k),$ $\mathbf{v}_f(k),$ and $\mathbf{w}_f(k)$, the state space form in \Cref{eq:2.12} can be rewritten as follows, which is widely known as the subspace form of the state space model.
\begin{align}
    \mathbf{y}_f(k) = \Gamma_f\mathbf{x}(k) + \mathbf{H}_f\mathbf{u}_f(k) - \mathbf{H}_f\mathbf{w}_f(k)\ + \mathbf{G}_f\mathbf{p}_f(k) + \mathbf{v}_f(k) \label{eq:2.15}
\end{align}
Here,
\begin{equation}
    \label{eq:2.16}
    \mathbf{\Gamma}_f = 
        \begin{bmatrix}
            \mathbf{C} \\ \mathbf{CA} \\ \vdots \\ \mathbf{CA}^{f-1} 
        \end{bmatrix} \in \mathbb{R}^{mf\ \times\ \eta}
\end{equation}
is the extended observability matrix with rank $\eta$. 
\begin{equation}
    \label{eq:2.17}
    \mathbf{H}_f = 
        \begin{bmatrix}
            \mathbf{D} & \mathbf{0} & \ldots & \mathbf{0} \\
            \mathbf{CB} & \mathbf{D} & \ldots & \mathbf{0} \\
            \vdots & \vdots & \ddots & \vdots \\
            \mathbf{CA}^{f-2}\mathbf{B} & \mathbf{CA}^{f-3}\mathbf{B} & \ldots & \mathbf{D} 
        \end{bmatrix} \in \mathbb{R}^{mf\ \times\ \ell f}
\end{equation}
and 
\begin{equation}
    \label{eq:2.18}
    \mathbf{G}_f = 
        \begin{bmatrix}
            \mathbf{0} & \mathbf{0} & \ldots & \mathbf{0} \\
            \mathbf{C} & \mathbf{0} & \ldots & \mathbf{0} \\
            \vdots & \vdots & \ddots & \vdots \\
            \mathbf{CA}^{f-2} & \mathbf{CA}^{f-2} & \ldots & \mathbf{0}
        \end{bmatrix} \in \mathbb{R}^{mf\ \times\ \eta f}
\end{equation}
are two block Toeplitz matrices. Introducing the lagged data vector $\mathbf{z}_f(k)$, which comprise the lagged output and input variable measurements as:
\begin{equation}
    \label{eq:2.19}
    \mathbf{z}_f(k) = 
        \begin{bmatrix}
            \mathbf{y}_f(k) \\ \mathbf{u}_f(k) 
        \end{bmatrix} \in \mathbb{R}^{(m+\ell)f}
\end{equation}
Therefore, the lagged data matrix can be defined as:
\begin{equation}
    \label{eq:2.20}
    \mathbf{Z}_f = \begin{bmatrix} \mathbf{z}_f(1) & \mathbf{z}_f(2) & \ldots & \mathbf{z}_f(N-f+1) \end{bmatrix}^\intercal 
\end{equation}
which is $(N-f+1) \times (m+\ell)f$, and can be used to rewrite the \Cref{eq:2.15} as $\begin{bmatrix} \mathbf{I}\ | -\mathbf{H}_f \end{bmatrix} \mathbf{Z}_f^\intercal = \mathbf{\Gamma}_f\mathbf{x}_f + \mathbf{E}_f^\intercal$, where $\mathbf{E}_f$ is constructed from $\mathbf{e}_f(k) = \mathbf{v}_f(k) - \mathbf{H}_f\mathbf{w}_f(k) + \mathbf{G}_f\mathbf{p}_f(k)$ likewise $\mathbf{Z}_f$ is constructed from $\mathbf{z}_f(k)$. Now, pre-multiplying both sides with $(mf-\eta)\ \times\ mf$ matrix $\left( \mathbf{\Gamma}_f^{\perp} \right)^\intercal$, orthogonal to $\mathbf{\Gamma}_f$, we get:
\begin{equation}
    \label{eq:2.21}
    \left( \mathbf{\Gamma}_f^{\perp} \right)^\intercal \begin{bmatrix} \mathbf{I}\ | -\mathbf{H}_f \end{bmatrix} \mathbf{Z}_f^\intercal = \left( \mathbf{\Gamma}_f^{\perp} \right)^\intercal \mathbf{E}_f^\intercal
\end{equation}
which, in absence of both the process and measurement noises, takes the following form:
\begin{equation}
    \label{eq:2.22}
    \left( \mathbf{\Gamma}_f^{\perp} \right)^\intercal \begin{bmatrix} \mathbf{I}\ | -\mathbf{H}_f \end{bmatrix} \mathbf{Z}_f^{*\intercal} = \mathbf{0}_{(mf-\eta)\times(N-f+1)}
\end{equation}
where, $\mathbf{Z}^*_f$ is lagged data matrix with the true values of lagged output and input variables. The measurement model for the noisy case can be written as follows:
\begin{equation}
    \label{eq:2.23}
    \mathbf{z}_f(k) = \mathbf{z}_f^*(k) + \begin{bmatrix} \mathbf{v}_f(k) + \mathbf{G}_f\mathbf{p}_f(k) \\ \mathbf{w}_f(k) \end{bmatrix}
\end{equation}

\Cref{eq:2.22,eq:2.23} are similar to the static model given by \Cref{eq:2.1}, where the equivalent constraint model in the dynamic case will take the form:
\begin{equation}
    \label{eq:2.24}
    \mathbf{A}_d = \left( \mathbf{\Gamma}_f^{\perp} \right)^\intercal \begin{bmatrix} \mathbf{I}\ | -\mathbf{H}_f \end{bmatrix}
\end{equation}
which can be used to extract the system matrices $\mathbf{A,B,C,}$ and $\mathbf{D}$ as discussed in \Cref{app:est}. It may also be noted that system matrices can only be estimated up to a similarity transformation. For instance, if $\mathbf{x}_t =  \mathbf{Tx}$ denotes the transformed states using a non-singular matrix $\mathbf{T}$, the state space model in \Cref{eq:2.12} can be equivalently written as:
\begin{subequations}
    \label{eq:2.25}
    \begin{align}
        &\mathbf{x}_t(k+1) = \mathbf{TAT}^{-1}\mathbf{x}_t(k) + \mathbf{TBu}^*(k) + \mathbf{Tp}(k) \\
        &\mathbf{y}^*(k) = \mathbf{CT}^{-1}\mathbf{x}_t(k) + \mathbf{Du}^*(k) 
    \end{align}
\end{subequations}
Consequently, the process noise covariance matrix of the equivalent state space model gets modified, while the measurement noise variances $\mathbf{\Sigma}_w, \mathbf{\Sigma}_v$ remain unchanged. This property is exploited for choosing appropriate state space representation to eliminate the need to estimate the process noise variances as described below. The noise covariance matrix of errors in lagged measurements $\mathbf{z}_f(k)$ is given by:
\begin{equation}
    \label{eq:2.26}
    \mathbf{\Sigma}_{\mathbf{e}f} = 
        \begin{bmatrix}
            \mathbf{G}_f\mathbf{\Sigma}_{pf}\mathbf{G}_f^\intercal + \mathbf{\Sigma}_{vf} & \mathbf{0} \\
            \mathbf{0} & \mathbf{\Sigma}_{wf}
        \end{bmatrix} \in \mathbb{R}^{(m+\ell)f \times (m+\ell)f}
\end{equation}
where,
\begin{subequations}
    \label{eq:2.27}
    \begin{align}
        &\mathbf{\Sigma}_{pf} = \mathbf{I}_f \otimes \mathbf{\Sigma}_p \in \mathbb{R}^{\eta f \times \eta f}; \label{eq:2.27a} \\
        &\mathbf{\Sigma}_{vf} = \mathbf{I}_f \otimes \mathbf{\Sigma}_v \in \mathbb{R}^{mf \times mf}; \label{eq:2.27b} \\
        &\mathbf{\Sigma}_{wf} = \mathbf{I}_f \otimes \mathbf{\Sigma}_w \in \mathbb{R}^{\ell f \times \ell f} \label{eq:2.27c}
    \end{align}
\end{subequations}

Due to the transformation of the system matrices and states, the estimated transformed noise covariance matrix takes the form $\mathbf{\hat{\Sigma}}_p = \mathbf{T\Sigma}_p\mathbf{T}^\intercal$. Assuming that the process noise is full rank, there exists a $\mathbf{T} = \mathbf{\Sigma}_p^{-1/2}$ such that $\mathbf{\hat{\Sigma}}_p$ is an identity matrix. Therefore, during the estimation phase, by assigning $\mathbf{\Sigma}_{pf}$ as identity matrix, the estimated lagged noise covariance matrix $\mathbf{\hat{\Sigma}}_{\mathbf{e}f}$ becomes:
\begin{equation}
    \label{eq:2.28}
    \mathbf{\hat{\Sigma}}_{\mathbf{e}f} = 
        \begin{bmatrix}
            \mathbf{\hat{G}}_f\mathbf{\hat{G}}_f^\intercal + \mathbf{\hat{\Sigma}}_{vf} & \mathbf{0} \\
            \mathbf{0} & \mathbf{\hat{\Sigma}}_{wf}
        \end{bmatrix}
\end{equation}
which also couples the estimation of system matrices $(\mathbf{A,C})$ since $\mathbf{G}_f$ is dependent on them. As a consequence of the mapping of this dynamic problem to an equivalent static problem, IPCA is used to obtain $\mathbf{A}_d$ (up to a rotation) and $\mathbf{\Sigma}_{\mathbf{e}f}$ simultaneously by iterating between two sub-steps. In the first sub-step, PCA is performed on the scaled lagged data matrix $\mathbf{Z_S}_f$ to obtain an estimate of the constraint matrix $\mathbf{\hat{A}}_d$, for a given estimate of $\mathbf{\hat{\Sigma}}_{\mathbf{e}f}$. In the second sub-step, the estimate of $\mathbf{\hat{\Sigma}}_{\mathbf{e}f}$ is updated by minimizing the following objective function:
\begin{align}
    \mathbf{\hat{\Sigma}}_{\mathbf{e}f}^{(i)} = \underset{\mathbf{\hat{\Sigma}}_{\mathbf{e}f}^{(i-1)}}{\mathrm{arg\ min}} \Biggl[& (N-f+1)\ \mathrm{log}\left| \mathbf{\hat{A}}_d^{(i)} \mathbf{\hat{\Sigma}}_{\mathbf{e}f}^{(i-1)} \left( \mathbf{\hat{A}}_d^{(i)} \right)^\intercal \right|\ + \nonumber \\ &\sum_{k=1}^{N-f+1} \left( \mathbf{r}^{(i)}(k) \right)^\intercal \left( \mathbf{\hat{A}}_d^{(i)} \mathbf{\hat{\Sigma}}_{\mathbf{e}f}^{(i-1)} \left(\mathbf{\hat{A}}_d^{(i)\intercal}\right) \right)^{-1} \mathbf{r}^{(i)}(k) \Biggr] \label{eq:2.29}
\end{align}
where, the superscript $(\cdot)^{(i)}$ denotes the iteration number. This incorporates few changes as compared to the objective function for the static case, described in \Cref{eq:2.8}. Due to stacking of lagged measurements, the number of observations is reduced to $(N-f+1)$. Further modification takes place in the identifiability constraint (Eq. \ref{eq:2.7}) to ensure that the noise variances are uniquely estimated along with $\mathbf{A}_d$ up to a rotation, which states that, given sufficient number of samples, it is always possible to choose the value of $f$ such that $(mf-\eta)(mf-\eta+1)/2 \geq (m+\ell)$. This leads to the following lower limit on number of constraints $d_{min}$:
\begin{equation}
    \label{eq:2.30}
    d_{min} (d_{min} + 1) \geq 2(m + \ell)
\end{equation}
since, $(m+\ell)$ is the number of decision variables in the above optimization problem, corresponding to which the measurement noise variances need to be estimated. The number of constraints is estimated by performing a hypothesis test with $d_{guess}$ starting from $d_{max} = mf-1$ up to $d_{min}$, until $\hat{d} = d_{guess}$. From the estimated number of constraints, the order of the system $\hat{\eta}$ is estimated as:
\begin{equation}
    \label{eq:2.31}
    \hat{\eta} = mf - \hat{d}
\end{equation}
The overall SMI-IPCA algorithm is summarized in \Cref{app:algo} for completeness.

\section{Proposed RSMI-IPCA methodology} \label{sec:rsmi-ipca}
In this section, we discuss the core methodological contribution of this work, that extends SMI-IPCA to handle the time-varying dynamical processes, by combining its mathematical rigor with the conceptual ideas of RIPCA methodology. We divide the complete procedure into two following parts, which explains the recursive re-estimation of the constraint model and the noise variances, followed by the estimation of the state-space model matrices from the updated estimates.

\subsection{Recursive update} \label{subsec:rsmi-ipca.step1}
The RIPCA algorithm, as discussed in \Cref{subsec:ripca}, employs a recursive covariance updating mechanism wherein newly acquired data matrix (at the $k$'th instant) is directly utilized. Similar strategy is systematically extended to the SMI-IPCA framework to address time-varying dynamical processes, thereby enabling adaptive parameter estimation in subspace-based EIV modeling. To effectively capture the temporal dependencies inherent in dynamic systems, the algorithm employs a lagged data matrix construction with stacking length $f$, a methodological approach established in \Cref{subsec:smiipca}. At each sampling instant, as we receive a new data vector, we only require measurements corresponding to the preceding $f-1$ time instances along with the sample covariance matrix of lagged measurements at the preceding time instant. 

Therefore, given the lagged sample covariance matrix $\mathbf{S}_{\mathbf{Z}_f,k-1}$ at $(k-1)$'th instant, the recursive update rule to obtain the current estimate of the lagged sample covariance matrix $\mathbf{S}_{\mathbf{Z}_f,k}$ follows from \Cref{eq:2.4} by substituting the data matrix with the appropriately constructed lagged data vector $\mathbf{z}_{f}(k)$ as:
\begin{equation}
    \label{eq:3.1}
    \mathbf{S}_{\mathbf{Z}_{f},k} = \frac{N_{k-1}}{N_k} \mathbf{S}_{\mathbf{Z}_{f},k-1} + \frac{1}{N_k} \mathbf{z}_{f}(k) \mathbf{z}_{f}(k)^\intercal
\end{equation}

In each recursive re-estimation step, analogous to the SMI-IPCA algorithm, the core idea is to iterate between the estimations of constraint model and noise covariance matrices, respectively. If an estimate of the lagged noise covariance matrix $\mathbf{\hat{\Sigma}}_{\mathbf{e}f,k-1}$ is known at time $k$, then the scaled and lagged sample covariance matrix can be computed by following the \Cref{eq:2.9} as:
\begin{subequations}
    \label{eq:3.2}
    \begin{align}
        &\mathbf{\hat{\Sigma}}_{\mathbf{e}f,k-1} = \mathbf{\hat{L}}_{\mathbf{e}f,k} \times \mathbf{\hat{L}}_{\mathbf{e}f,k}^\intercal \\
        &\implies \mathbf{S}_{\mathbf{Z_S}_{f},k} = \mathbf{\hat{L}}_{\mathbf{e}f,k}^{-\intercal} \left( \mathbf{S}_{\mathbf{Z}_{f},k} \right) \mathbf{\hat{L}}_{\mathbf{e}f,k}^{-1}
    \end{align}
\end{subequations}
on which, eigenvalue decomposition is performed to obtain an estimate of the constraint model matrix $\mathbf{\hat{A}}_{d,k}$ at $k$'th instant. Subsequently in the second step, the lagged noise covariance matrix is updated using the constraint model estimate by solving the optimization problem defined by \Cref{eq:2.29}. However, the objective function requires subtle modification as also described in \Cref{eq:2.10,eq:2.11}, in order to eliminate the need of storing complete historical data. The newly estimated $\mathbf{\hat{A}}_{d,k}$ matrix is further utilized to obtain updated estimates of $(\mathbf{\hat{A}},\mathbf{\hat{C}})$ model matrices, followed by computing the $\mathbf{\hat{G}}_f$ matrix as described in \Cref{app:est}. Therefore, a subtly updated version of the previously estimated lagged noise covariance matrix $\mathbf{\hat{\Sigma}}_{\mathbf{e}f,k-1}$ is used in the following modified objective function:
\begin{align}
    \mathbf{\hat{\Sigma}}_{\mathbf{e}f,k} = \mathrm{arg}\ \underset{\mathbf{\hat{\Sigma}}_{\mathbf{e}f,k-1}}{\mathrm{min}} \Biggl[& (N_k-f+1)\ \mathrm{log}\left| \mathbf{\hat{A}}_{d,k} \mathbf{\hat{\Sigma}}_{\mathbf{e}f,k-1} \mathbf{\hat{A}}_{d,k}^\intercal \right|\ + \nonumber \\ &(N_k-f+1) \mathrm{Tr}\left( \left( \mathbf{\hat{A}}_{d,k} \mathbf{\hat{\Sigma}}_{\mathbf{e}f,k-1} \mathbf{\hat{A}}_{d,k}^\intercal \right)^{-1} \mathbf{\hat{A}}_{d,k} \mathbf{S}_{\mathbf{Z}_{f},k} \mathbf{\hat{A}}_{d,k}^\intercal \right) \Biggr] \label{eq:3.3}
\end{align}

It is evident from \Cref{eq:3.3}, that the objective function of the above optimization problem requires only historical data for the preceding $f$ time instants ($\mathbf{S}_{\mathbf{Z}_{f},k}$) for computing its covariance matrix, and not the entire historical data. The solution of the optimization problem provides new estimates of the noise variances.  Iterating between the two steps of the algorithm until convergence provides updated estimates of the order, the state space model matrices, and input and output error variances. The recursive SMI-IPCA algorithm is referred to as RSMI-IPCA.

\subsection{Estimation of system matrices} \label{subsec:rsmi-ipca.step2}
The recursive re-estimation requires the determination of $(\mathbf{\hat{A}}, \mathbf{\hat{C}})$ model matrices at each iteration of the RSMI-IPCA method, because the noise covariance matrix is dependent on these state space model matrices as discussed in \Cref{subsec:rsmi-ipca.step1}. The reader is referred to \Cref{app:est} for the detailed estimation procedure of the model matrices. Since the state space estimated using RSMI-IPCA is a specific convenient form chosen to make the process noise covariance matrix to be an identity matrix (thus eliminating the need to estimate it), comparison of the estimated model with any other form of the state space model can be made only using system invariant parameters such as the poles and zeros. A detailed description of the proposed RSMI-IPCA algorithm is provided in \Cref{algo:1}, for completeness.

\begin{algorithm}[!ht]
    \caption{The RSMI-IPCA algorithm for tracking time-varying dynamic processes}
    \footnotesize
    \KwIn{The data $\mathbf{z}_{f}(k),$ and lag $f$}
    \KwRequire{From $(k-1)$'th instant, $\mathbf{S}_{\mathbf{Z}_f,k-1},$ and $N_{k-1}$, estimates of $\mathbf{\hat{\Sigma}}_{v,k-1},$ $\mathbf{\hat{\Sigma}}_{w,k-1},$ $\mathbf{\hat{A}}_{k-1},$ and $\mathbf{\hat{C}}_{k-1}$}
    \KwParam{Maximum iterations $i_{max} >0,$ stopping tolerance $\varepsilon_\lambda > 0$}
    \KwOut{Updated estimates of $\mathbf{\hat{\Sigma}}_{v,k},$ $\mathbf{\hat{\Sigma}}_{w,k}$, model matrices $\left(\mathbf{\hat{A}}_{k}, \mathbf{\hat{C}}_{k} \right)$, and process order $\hat{\eta}_k$}

        \nl\label{alg1:st1}Compute $N_k = N_{k-1} + 1$\;
        
        \nl\label{alg1:st2}Initialize $i=1,$ $\lambda^{(0)} =0,$ and $\lambda^{(i)} = 1$. Set $\mathbf{\hat{\Sigma}}_{v,k}^{(i)} \leftarrow \mathbf{\hat{\Sigma}}_{v,k-1},$ $\mathbf{\hat{\Sigma}}_{w,k}^{(i)} \leftarrow \mathbf{\hat{\Sigma}}_{w,k-1},$ $\mathbf{\hat{A}}_{k}^{(i)} \leftarrow \mathbf{\hat{A}}_{k-1},$ and $\mathbf{\hat{C}}_{k}^{(i)} \leftarrow \mathbf{\hat{C}}_{k-1}$\;

        \nl\label{alg1:st3}Compute $\mathbf{\hat{G}}_f^{(i)}$ using $\mathbf{\hat{A}}_k^{(i)},\mathbf{\hat{C}}_k^{(i)}$ from \Cref{eq:2.18}\;

        \nl\label{algo:st4}Obtain $\mathbf{S}_{\mathbf{Z}_{f},k}$ using $\mathbf{S}_{\mathbf{Z}_{f},k-1}$ and $\mathbf{z}_{f}(k)$ based on \Cref{eq:3.1}\;

        \nl\label{alg1:st5}\While{$i \leq i_{max}$ $\mathbf{and}$ $\left|\left(\lambda^{(i)} - \lambda^{(i-1)}\right) / \lambda^{(i-1)}\right| > \varepsilon_\lambda$} {
        
            \nl\label{alg1:st6}Construct $\mathbf{\hat{\Sigma}}_{\mathbf{e}f, k-1}^{(i)}$ from $\mathbf{\hat{G}}_f^{(i)}, \mathbf{\hat{\Sigma}}_{v,k}^{(i)},$ and $\mathbf{\hat{\Sigma}}_{w,k}^{(i)}$ using \Cref{eq:2.28}\;

            \nl\label{alg1:st7}Scale $\mathbf{S}_{\mathbf{Z}_{f},k}$ using $\mathbf{\hat{L}}_{\mathbf{e}f,k}^{(i)}$, the Cholesky factor of $\mathbf{\hat{\Sigma}}_{\mathbf{e}f, k-1}^{(i)}$ as shown in \Cref{eq:3.2} to obtain $\mathbf{S}_{\mathbf{Z_S}_{f},k}$\;

            \nl\label{alg1:st8}$\left[\mathbf{\hat{V}\ \mathbf{\hat{D}}} \right] = \mathrm{eig}\left( \mathbf{S}_{\mathbf{Z_S}_{f},k} \right)$. Obtain $\hat{d}$ by employing hypothesis test on $\mathbf{S}_\lambda$ as detailed in \cite{Keerthan:2023}\;

            \nl\label{alg1:st9}Estimate constraint matrix $\mathbf{\hat{A}}_{\hat d,k}^{(i+1)} = \mathbf{\hat{V}}_{\hat{d}}^\intercal \times \mathbf{\hat{L}}_{\mathbf{e}f,k}^{(i)-1}$. Estimate the process order as $\hat{\eta}_k = mf - \hat{d}$\;

            \nl\label{alg1:st10}Compute $\mathbf{\hat{\Gamma}}_f$ from $\mathbf{\hat{A}}_{\hat d,k}^{(i+1)}$ using \Cref{eq:app.B1}\;

            \nl\label{alg1:st11}Compute $\mathbf{\hat{A}}_k^{(i+1)}, \mathbf{\hat{C}}_k^{(i+1)}$ from $\mathbf{\hat{\Gamma}}_f$ using \Cref{eq:app.B3a,eq:app.B3b}\;

            \nl\label{alg1:st12}Compute $\mathbf{\hat{G}}_f^{(i+1)}$ using $\mathbf{\hat{A}}_k^{(i+1)},\mathbf{\hat{C}}_k^{(i+1)}$ from \Cref{eq:2.18}\;

            \nl\label{alg1:st13}Solve \Cref{eq:3.3} using $\mathbf{S}_{\mathbf{Z}_{f},k}, \mathbf{\hat{A}}_{\hat d,k}^{(i+1)}, \mathbf{\hat{G}}_f^{(i+1)}, \mathbf{\hat{\Sigma}}_{v,k}^{(i)},$ and $\mathbf{\hat{\Sigma}}_{w,k}^{(i)}$ to obtain new estimates of $\mathbf{\hat{\Sigma}}_{v,k}^{(i+1)}$ and $\mathbf{\hat{\Sigma}}_{w,k}^{(i+1)}$\;

            \nl\label{alg1:st14}Set $\lambda^{(i-1)} \leftarrow \lambda^{(i)},$ $i \leftarrow i+1$\;
            
            \nl\label{alg1:st15}Compute $\lambda^{(i)}$ as the trace of the $\hat{d}$ smallest eigenvalues contained in $\mathbf{\hat{D}}$\;
        }
        \nl\label{alg1:st16}Set $\mathbf{\hat{\Sigma}}_{v,k} \leftarrow \mathbf{\hat{\Sigma}}_{v,k}^{(i+1)},$ $\mathbf{\hat{\Sigma}}_{w,k} \leftarrow \mathbf{\hat{\Sigma}}_{w,k}^{(i+1)},$ $\mathbf{\hat{A}}_{k} \leftarrow \mathbf{\hat{A}}_{k}^{(i+1)},$ and $\mathbf{\hat{C}}_{k} \leftarrow \mathbf{\hat{C}}_{k}^{(i+1)}$\;
        
        \nl\label{alg1:st17}Compute $\mathbf{\hat{H}}_{f1}$ from $\mathbf{\hat{\Gamma}}_f$ and $\mathbf{\hat{A}}_{d,k}^{(i+1)}$ using \Cref{eq:app.B6} \Comment*[r]{Optional system matrix identification step}

        \nl\label{alg1:st18}Estimate $\mathbf{\hat{B}}_{k}, \mathbf{\hat{D}}_{k}$ from $\mathbf{\hat{\Gamma}}_f, \mathbf{\hat{H}}_{f1}$ from \Cref{eq:app.B7a,eq:app.B7b}\;
    
    \label{algo:1}
\end{algorithm}

\section{Case studies} \label{sec:simulation}
We discuss the efficacy and functioning of the proposed RSMI-IPCA algorithm using three case studies on two systems. We consider three practical application scenarios where online monitoring of the dynamic system becomes crucial. In the first example, we consider a benchmark fourth-order dynamic process drawn from \cite{Wang:2002} to simulate gradual degradation in sensor precision. We further demonstrate the ability of this algorithm in tracking the time-varying model coefficients of a non-interacting two-tank system in series \cite{Tangirala:2015} under the influence of time-varying operational conditions. The same system is further used in the third study, where we demonstrate the algorithm's capability to track the changes in model structure and process order.

We have considered a sampling instance of $1$s and for each new sample received, we have applied the RSMI-IPCA algorithm. In order to evaluate its performance, we consider the following three metrics computed after each sample. 
\begin{enumerate}
    \item The absolute difference between the estimated and true process order is given by $\left| \hat{\eta} - \eta \right|$.
    
    \item The sum of the relative differences between the estimated and true noise standard deviations (SD):
    \begin{equation}
        \label{eq:4.1}
        \sum_{i=1}^{m} \left| \frac{\hat{\sigma}_{y_i,k} - \sigma_{y_i}}{\sigma_{y_i}} \right| + \sum_{j=1}^{\ell} \left| \frac{\hat{\sigma}_{u_j,k} - \sigma_{u_j}}{\sigma_{u_j}} \right|
    \end{equation}
    where, $\hat{\sigma}_{y_i,k}$ and $\hat{\sigma}_{u_j,k}$ are the estimated noise SD at $k$'th instant of $i$'th output and $j$'th input variables, respectively. $\sigma_{y_i}$ and $\sigma_{u_j}$ are the corresponding true noise SDs.
    
    \item Specifically, for the two-tank system, we use this additional metric, which is the Frobenius norm of the difference between the estimated and true system poles, defined as $\|\mathbf{\hat{P}} - \mathbf{P}\|_F$. While running the experiment, when the estimated process order is not equal to the true process order, either of the pole vectors is padded with zeros to make them dimensionally compatible.
\end{enumerate}

\subsection{System 1: A fourth-order dynamic process} \label{subsec:system1}
Consider the fourth-order $2\times 2$ dynamic process, drawn from \cite{Wang:2002} as described below:
\begin{subequations}
    \label{eq:4.3}
    \begin{align}
        \mathbf{x}(k+1) =& 
            \begin{bmatrix}
                0.67 & 0.67 & 0 & 0 \\
                -0.67 & 0.67 & 0 & 0 \\
                0 & 0 & -0.67 & -0.67 \\
                0 & 0 & 0.67 & -0.67
            \end{bmatrix} \mathbf{x}(k) + \begin{bmatrix}
                0.6598 & -0.5256 \\
                1.9698 & 0.4845 \\
                4.3171 & -0.4879 \\
                -2.6436 & -0.3416
            \end{bmatrix} \mathbf{u}^*(k) \\
        \mathbf{y}^*(k) =& 
            \begin{bmatrix}
                -0.5749 & 1.0751 & -0.5225 & 0.1830 \\
                2.4027 & 0.7543 & -0.2159 & 0.0982
            \end{bmatrix} \mathbf{x}(k) + \begin{bmatrix}
                -0.7139 & -0.1174 \\ 0.3131 & -0.2876
            \end{bmatrix} \mathbf{u}^*(k)
    \end{align}
\end{subequations}
The noise-free inputs $\mathbf{u}^*$ are chosen to be full-band random binary signals (RBS) of length $N=4095$, which are further scaled to make the variance of the inputs equal to unity. Then, the output sequences $\mathbf{y}^*$ are generated using \Cref{eq:2.12}, by considering $\mathbf{p} \sim \mathcal{N}(\mathbf{0}, \mathbf{\Sigma}_p)$, where $\mathbf{\Sigma}_p$ is chosen to be diagonal with entries $0.49, 0.36, 0.64,$ and $0.25$, respectively. 

\subsubsection{Case study 1: Tracking sensor degradation} \label{subsubsec1:sensor}
\begin{figure}[!ht]
    \centering

    \begin{subfigure}[t]{0.48\columnwidth}
        \centering
        \includegraphics[width=1.0\columnwidth]{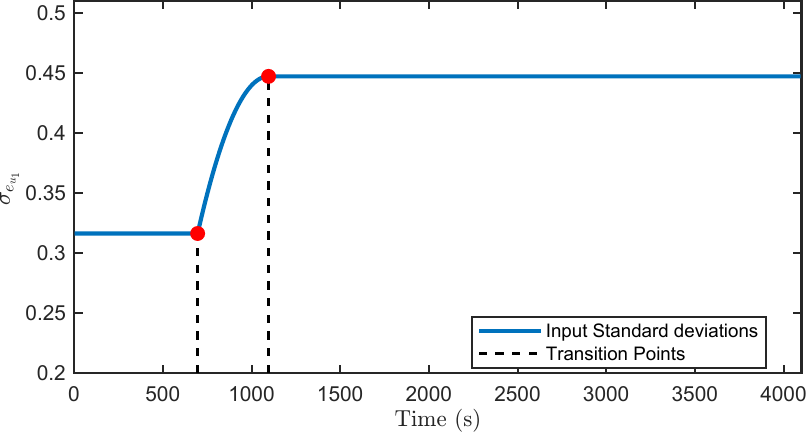}
        \subcaption[Short list entry]{Variation of the standard deviation corresponding to the first input variable's measurements}
        \label{Figure_1a}
    \end{subfigure}
    \hfill
    \begin{subfigure}[t]{0.48\columnwidth}
        \centering
        \includegraphics[width=1.0\columnwidth]{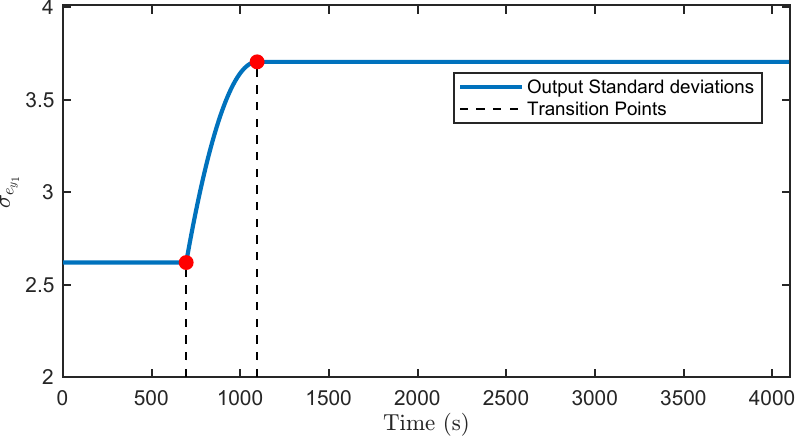}
        \subcaption[Short list entry]{Variation of the standard deviation corresponding to the first output variable's measurements}
        \label{Figure_1b}
    \end{subfigure}
    
    \caption{Simulation of gradual sensor degradation, i.e., increment in the noise standard deviations of input and output measurements in a quadratic manner from $695$'th instant to $1095$'th instant.}
    \label{Figure_1}
\end{figure}

In order to simulate the scenario of gradual degradation in sensor's precision, we corrupt the first $695$ samples of $\mathbf{u}^*$ and $\mathbf{y}^*$ using Gaussian noise sequences having fixed noise covariances $\mathbf{\Sigma}_v^{(1)} = \text{diag}\left(2.6192^2,\ 4.0644^2\right)$ and $\mathbf{\Sigma}_w^{(1)} = \text{diag}\left(0.3162^2,\ 0.3162^2\right)$, respectively such that the signal-to-noise ratio (SNR) is $10$. Thereafter, all the noise variances are gradually increased in a quadratic fashion up to the $1095$'th instant resulting in $\mathbf{\Sigma}_v^{(2)} = \text{diag}\left(3.7041^2,\ 5.7479^2\right)$ and $\mathbf{\Sigma}_w^{(2)} = \text{diag}\left(0.4472^2,\ 0.4472^2\right)$, which are kept fixed to corrupt the remaining sequence of samples. The new variances correspond to a SNR of $5$ and the variational trend of the standard deviations corresponding to the first input and output variables are reported in \Cref{Figure_1}. 

\begin{figure}[!ht]
    \centering

    \begin{subfigure}[t]{0.48\columnwidth}
        \centering
        \includegraphics[width=1.0\columnwidth]{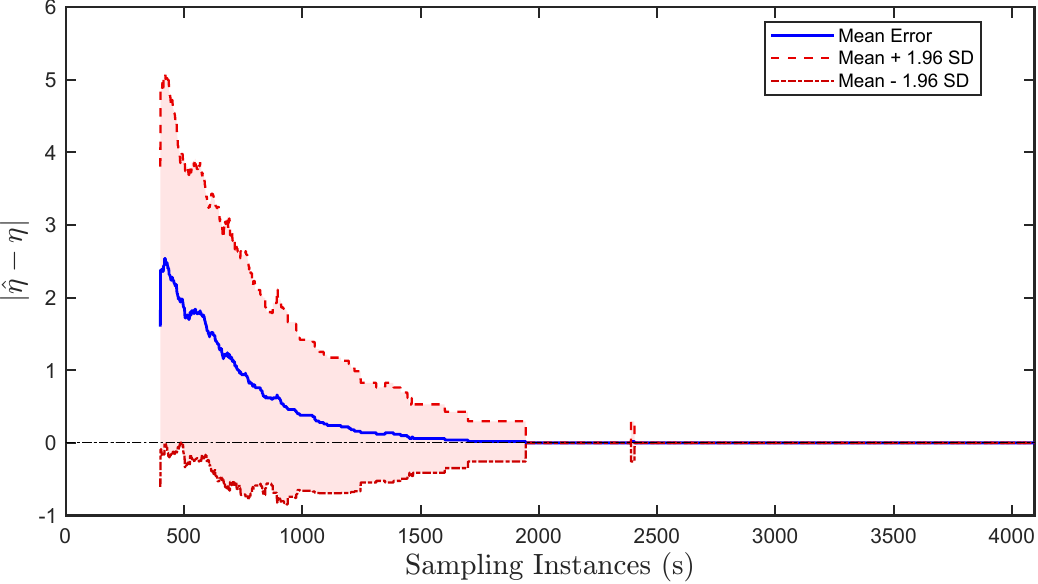}
        \subcaption[Short list entry]{Absolute difference between the estimated and true process orders of the fourth order system}
        \label{Figure_2a}
    \end{subfigure}
    \hfill
    \begin{subfigure}[t]{0.48\columnwidth}
        \centering
        \includegraphics[width=1.0\columnwidth]{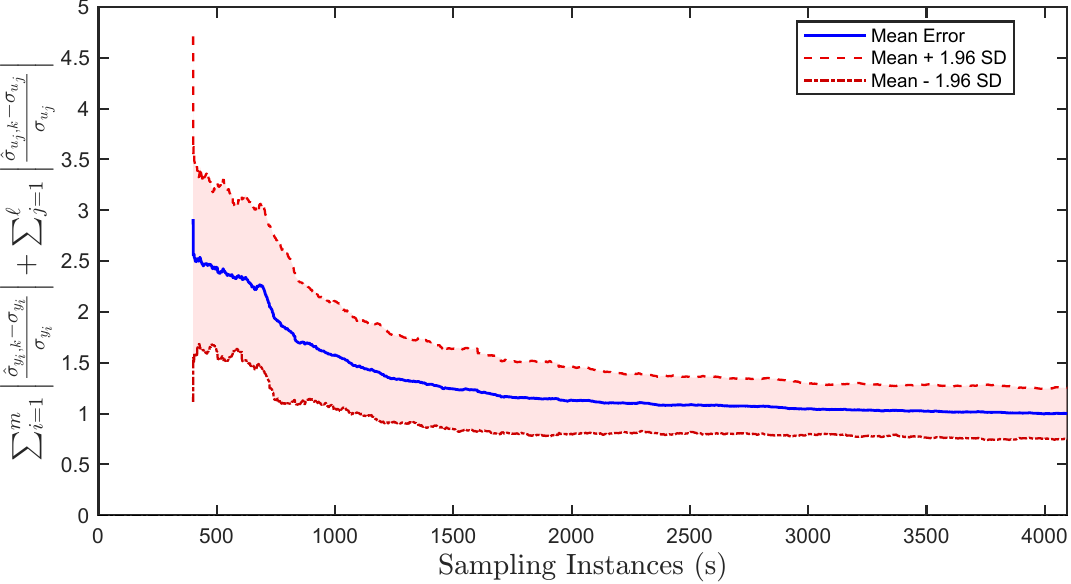}
        \subcaption[Short list entry]{Sum of the relative differences in the estimates of the noise standard deviations}
        \label{Figure_2b}
    \end{subfigure}
    
    \caption{Performance analysis of RSMI-IPCA for the LTI fourth-order system as described in \Cref{eq:4.3} under gradual sensor degradation.}
    \label{Figure_2}
\end{figure}

The initial estimates of the process order, noise variances and $(\mathbf{\hat{A}}, \mathbf{\hat{C}})$ model matrices, required to kick-start RSMI-IPCA as mentioned in \Cref{algo:1}, are obtained by applying the SMI-IPCA algorithm, taking the first $400$ samples and using a lag $f=6$. Hereafter, as new measurements are received, we apply RSMI-IPCA to recursively update these parameters and subsequently analyze the performance of this algorithm using the pre-defined performance metrics. We report the results by averaging over $50$ simulation trials with different noise realizations along with the respective $95\%$ confidence intervals of the first two metrics in \Cref{Figure_2}. The estimate of the process order converges to the true value, i.e., $\eta = 4$ within around $2000$ samples as observed from \Cref{Figure_2a}. Although after we introduce the changes in the noise variances, the assumption for the historical data to have the same noise covariance matrix gets violated, our proposed RSMI-IPCA is still able to maintain accurate tracking of the changes as we see in \Cref{Figure_2b}. This is due to re-estimating the noise variances at each step as new data is received. 

\begin{figure}[!htbp]
    \centering

    \begin{subfigure}[t]{0.48\columnwidth}
        \centering
        \includegraphics[width=1.0\columnwidth]{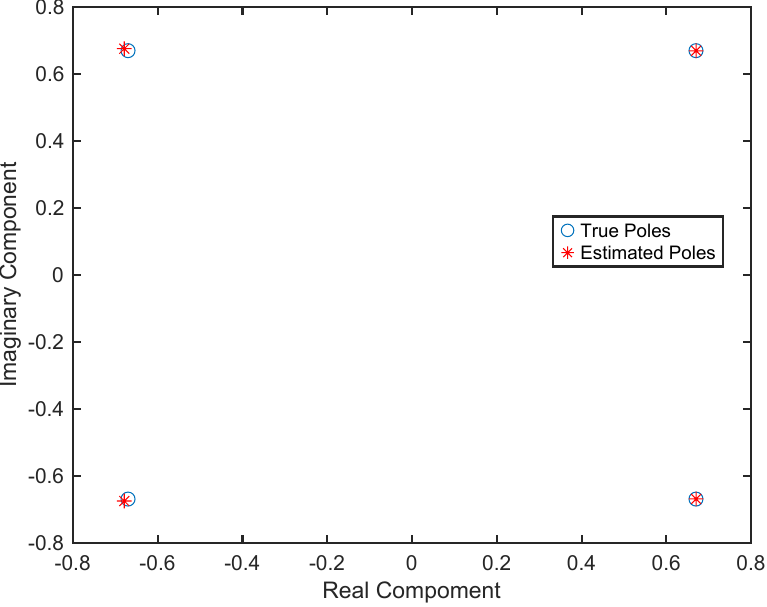}
        \subcaption[Short list entry]{Comparing the estimated and true poles}
        \label{Figure_3a}
    \end{subfigure}
    \hfill
    \begin{subfigure}[t]{0.48\columnwidth}
        \centering
        \includegraphics[width=1.0\columnwidth]{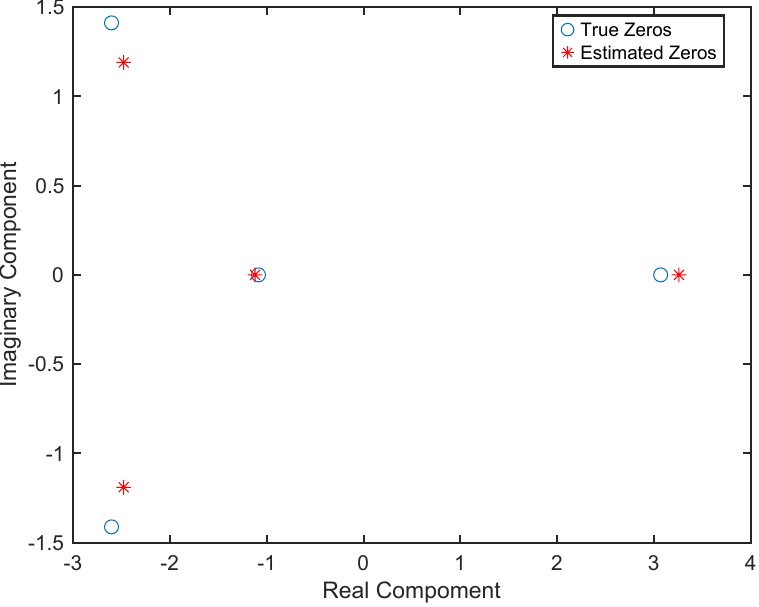}
        \subcaption[Short list entry]{Comparing the estimated and true zeros}
        \label{Figure_3b}
    \end{subfigure}
    
    \caption{The mean of the estimates of poles and zeros of the system defined in \Cref{eq:4.3} are observed to be pretty close to the respective true values.}
    \label{Figure_3}
\end{figure}

The final estimates of the measurement noise standard deviations are provided in \Cref{table:1}. We observe that the estimate of the $\hat{\sigma}_{e_{u_2}}$ corresponding to the second input variable is not very accurate whereas all other estimates are fairly close to their respective true values. As identified by Ramnath and Narasimhan~\cite{Keerthan:2023}, the estimates become more accurate with an appropriately larger choice of the value of lag $f$. Due to the presence of rotational ambiguity in the estimated constraint matrix, we cannot directly compare the estimated model matrices. However, we compare the poles and zeros of the true and estimated system, which can be seen in \Cref{Figure_3}. The estimates are averaged over the $50$ simulation trials and they come out to be pretty close to their respective true values. In \Cref{table:2}, we provide detailed statistics of the poles and zero estimates. The results indicate the unbiased nature of the estimated poles and zeros which contain their respective true values within their $95\%$ confidence interval.

\begin{table}[!ht]
\centering
\begin{minipage}{\columnwidth}
\centering
\begin{threeparttable}
\caption{Estimates of the noise SDs under sensor degradation.}
\label{table:1}
\renewcommand{\arraystretch}{1.25}
\begin{tabular}{c@{\hspace{3em}}c@{\hspace{3em}}c@{\hspace{2em}}c}
\toprule
$\sigma_i$ & True Values & $\hat{\sigma}_i$ & $1.96\times \sigma\left(\hat{\sigma}_i\right)$  \\
\toprule
$\sigma_{e_{y_1}}$ & $3.7041$ & $3.4571$ & $0.2593$ \\
$\sigma_{e_{y_2}}$ & $5.7479$ & $5.5107$ & $0.2620$ \\
$\sigma_{e_{u_1}}$ & $0.4472$ & $0.4543$ & $0.0451$ \\
$\sigma_{e_{u_2}}$ & $0.4472$ & $0.8259$ & $0.1952$ \\
\bottomrule
\end{tabular}
\begin{tablenotes}[flushleft]
\footnotesize
\item \hspace*{-0.1cm}Here, $\sigma\left( \hat{\sigma}_i \right)$ denotes the standard deviation of the estimated noise standard deviation, obtained based on the $50$ simulation trials.
\end{tablenotes}
\end{threeparttable}
\end{minipage}
\end{table}

\begin{table}[!ht]
\centering
\begin{minipage}{\columnwidth}
\centering
\begin{threeparttable}
\caption{Estimates of poles and zeros of the system in \Cref{eq:4.3}.}
\label{table:2}
\renewcommand{\arraystretch}{1.25}
\begin{tabular}{cccc}
\toprule
True poles $(p)$ & $\mu\left( \hat{p} \right)$ & $1.96 \times \sigma\left( \Re(\hat{p}) \right)$ & $1.96\times \sigma\left( \Im(\hat{p}) \right)$ \\
\midrule
$-0.670 - 0.670i$ & $-0.679 - 0.669i$ & $0.011$ & $0.005$ \\
$-0.670 + 0.670i$ & $-0.679 + 0.669i$ & $0.011$ & $0.005$ \\
$0.670 - 0.670i$ & $0.670 - 0.676i$ & $0.003$ & $0.011$ \\
$0.670 + 0.670i$ & $0.670 + 0.676i$ & $0.003$ & $0.011$ \\
\midrule
\midrule
True zeros $(z)$ & $\mu\left( \hat{z} \right)$ & $1.96\times \sigma\left( \Re(\hat{z}) \right)$ & $1.96\times \sigma\left( \Im(\hat{z}) \right)$ \\
\midrule
$-2.602 - 1.412i$ & $-2.477 - 1.190i$ & $0.745$ & $0.597$ \\
$-2.602 + 1.412i$ & $-2.477 + 1.190i$ & $0.745$ & $0.597$ \\
$-1.085$ & $-1.121$ & $0.109$ & -- \\
$3.067$ & $3.256$ & $0.656$ & -- \\
\bottomrule
\end{tabular}
\begin{tablenotes}[flushleft]
\footnotesize
\item \hspace*{-0.1cm}Here, we use $\mu(\cdot)$ and $\sigma(\cdot)$ to denote the mean and standard deviations, respectively, whereas $\Re(\cdot)$ and $\Im(\cdot)$ denote the real and imaginary components, respectively. 
\end{tablenotes}
\end{threeparttable}
\end{minipage}
\end{table}

\subsection{System 2: Non-interacting two-tank system} \label{subsec:system2}
\begin{figure}[!htbp]
    \centering
    \includegraphics[width=0.4\columnwidth]{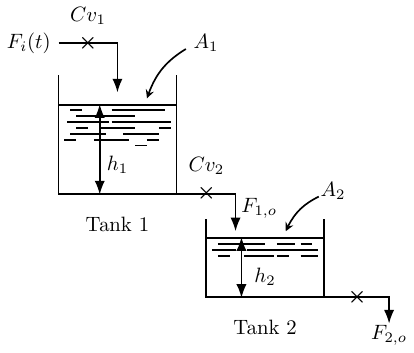}
    \caption{Schematic of the non-interacting two-tanks in series.}
    \label{Figure_4}
\end{figure}

In this case study, we demonstrate the ability of the proposed algorithm to track the changes in process operating conditions and model structure. For this purpose, we consider a system comprising two non-interacting tanks in series as shown in \Cref{Figure_4}. Based on the conservation of mass, the deterministic first-principles model of the liquid level system is \cite{Tangirala:2015}:
\begin{subequations}
    \label{eq:4.4}
    \begin{align}
        &\frac{dh_1(t)}{dt} = \frac{F_i(t)}{A_1} - \frac{Cv_1}{A_1}\sqrt{h_1(t)} \label{eq:4.4a} \\
        &\frac{dh_2(t)}{dt} = \frac{Cv_1}{A_2} \sqrt{h_1(t)} - \frac{Cv_2}{A_2}\sqrt{h_2(t)} \label{eq:4.4b}
    \end{align}
\end{subequations}
where, $F_i(t)$ is the input flow rate in the tank $1$, with liquid level $h_1(t)$ and cross-sectional area $A_1$. The liquid level in tank $2$ is $h_2(t)$, which has a cross-sectional area $A_2$. For both the following case studies, we fix the cross-sectional areas as $A_1=2.4,$ and $A_2=1.2$. $Cv_1,$ and $Cv_2$ are valve coefficients.

\subsubsection{Case study 2: Tracking changes in process operating conditions} \label{subsubsec2:process}
In this case study, we simulate the scenario of changing process operating conditions by varying the valve coefficients and demonstrate the applicability of RSMI-IPCA to track these changes.

\begin{figure}[!htbp]
    \centering

    \begin{subfigure}[t]{0.48\columnwidth}
        \centering
        \includegraphics[width=1.0\columnwidth]{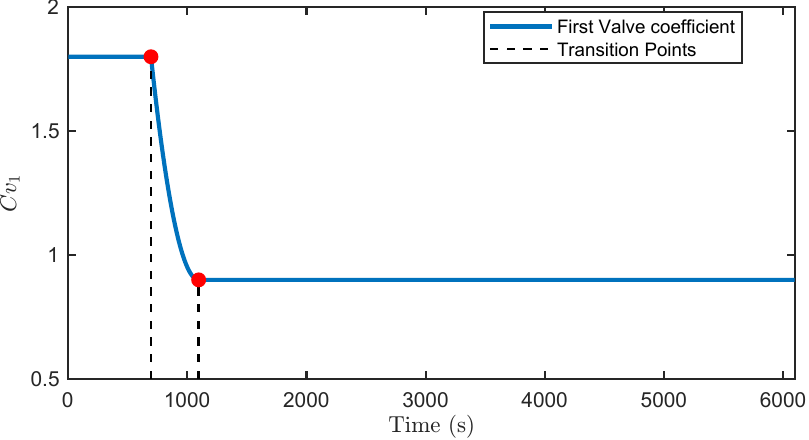}
        \subcaption[Short list entry]{Variation in the values of first valve coefficient}
        \label{Figure_5a}
    \end{subfigure}
    \hfill
    \begin{subfigure}[t]{0.48\columnwidth}
        \centering
        \includegraphics[width=1.0\columnwidth]{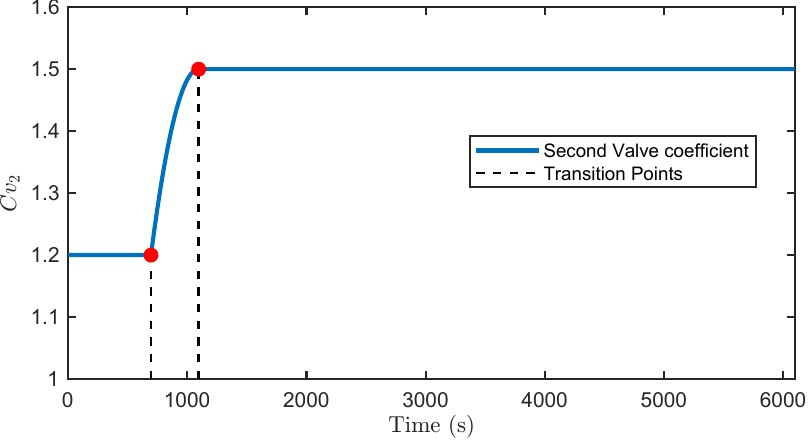}
        \subcaption[Short list entry]{Variation in the values of second valve coefficient}
        \label{Figure_5b}
    \end{subfigure}
    
    \caption{Simulation of variational process operational conditions, where the valve coefficients are changed gradually in a quadratic manner from $695$'th instant to $1095$'th instant.}
    \label{Figure_5}
\end{figure}

At first, the system is brought to steady state and thereafter excited with the designed input. The steady-state operating point of the input, $F_i(t)$ is chosen to be $2$ units and a full-band RBS of sample size $N=6095$, scaled with a factor $0.25$, is used to excite the steady-state value to generate the input sequence of length $N$ with a sampling rate of $1$s. Of specific interest to change process operating conditions, we generate the initial $695$ samples of the output $h_2(t)$ using the valve coefficient values $Cv_1^{(1)} = 1.8, Cv_2^{(1)} = 1.2$. For the next $400$ sampling instances, the valve coefficients are changed gradually, as reported in \Cref{Figure_5}. From $1095$'th instant onward, the valve coefficients are fixed to the updated values $Cv_1^{(2)} = 0.9$ and $Cv_2^{(2)} = 1.5$. $50\%$ reduction in $Cv_1$ can correspond to clogging effect in the pipe causing the inflow in the first tank, whereas $25\%$ increment in the value of $Cv_2$ can be caused by erosion effect from the pipe connected at the bottom of second tank. The order of the system can be verified to be $\eta = 2$ with respect to the input $F_i$ and output $h_2$. The main objective of this case study is to demonstrate the ability of RSMI-IPCA to track the changes in model parameters due to the changes in operating points. Therefore, we choose fixed standard deviations $\sigma_{F_i} = 0.1581$ and $\sigma_{h_2} = 0.1315$, corresponding to SNR = $10$, of the two Gaussian white noise sequences of length $N$ that corrupt the input and output measurements.

RSMI-IPCA is initialized by the estimates of process order, noise variances and state-space model matrices obtained by applying SMI-IPCA on the first $400$ samples using a lag $f=6$. Thereafter, as we receive subsequent samples at a sampling rate of $1$s, we apply RSMI-IPCA algorithm to update the process order, noise variances and the model matrices. Besides assessing the performance of this algorithm using the three performance metrics defined earlier, we also analyze the convergence of the initial system to the final updated system after the changes in the valve coefficients are introduced by tracking the poles\footnote{It is to note that the transfer function of this system has no zeros.}, estimated using the model matrices obtained from the updated constraint matrix at each step of applying this algorithm. All these results are averaged over $50$ simulation trials with different noise realizations and the mean value of the metrics along with their respective $95\%$ confidence band are provided in each figure for significance analysis of the estimates. 

\begin{figure}[!htb]
    \centering

    \begin{subfigure}[t]{0.48\columnwidth}
        \centering
        \includegraphics[width=1.0\columnwidth]{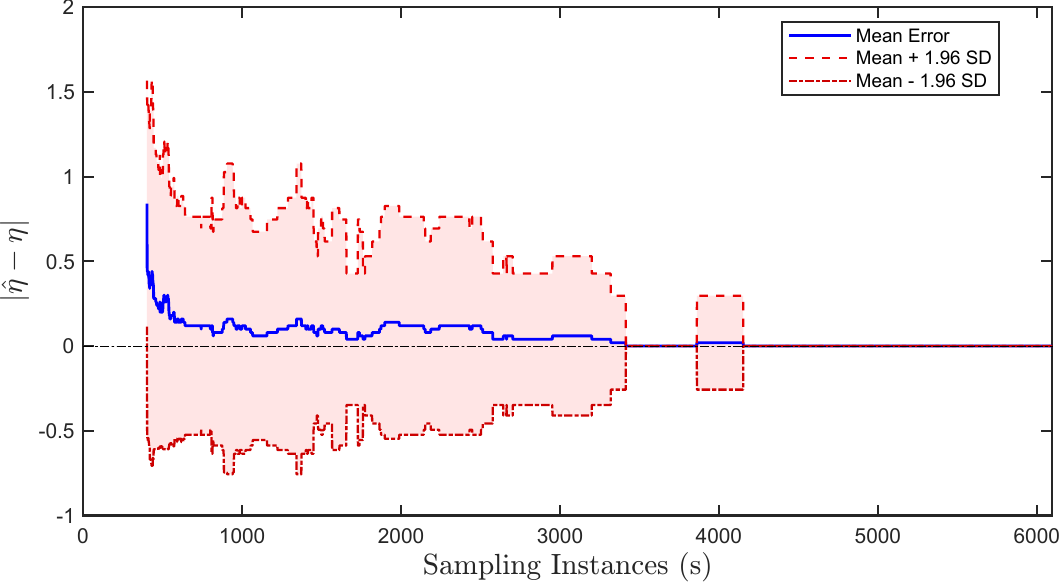}
        \subcaption[Short list entry]{The process order estimate converges gradually to its true value $\eta = 2$ around $3500$'th sampling instant}
        \label{Figure_6a}
    \end{subfigure}
    \hfill
    \begin{subfigure}[t]{0.48\columnwidth}
        \centering
        \includegraphics[width=1.0\columnwidth]{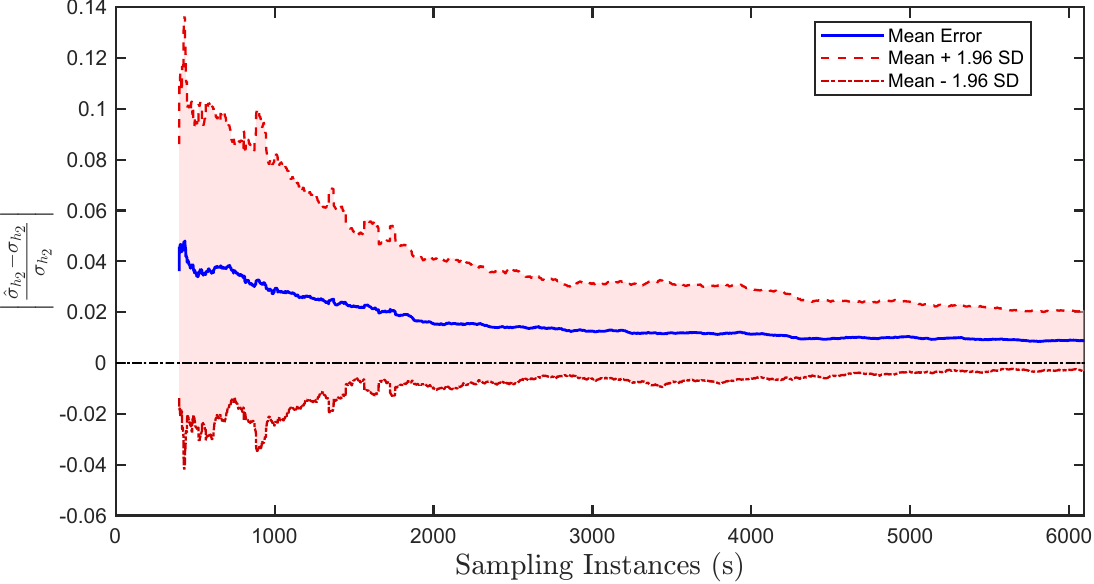}
        \subcaption[Short list entry]{The relative error in the estimate of the SD corresponding to output $h_2$ gradually reduces to zero}
        \label{Figure_6b}
    \end{subfigure}

    \vspace{0.8em}

    \begin{subfigure}[t]{0.48\columnwidth}
        \centering
        \includegraphics[width=1.0\columnwidth]{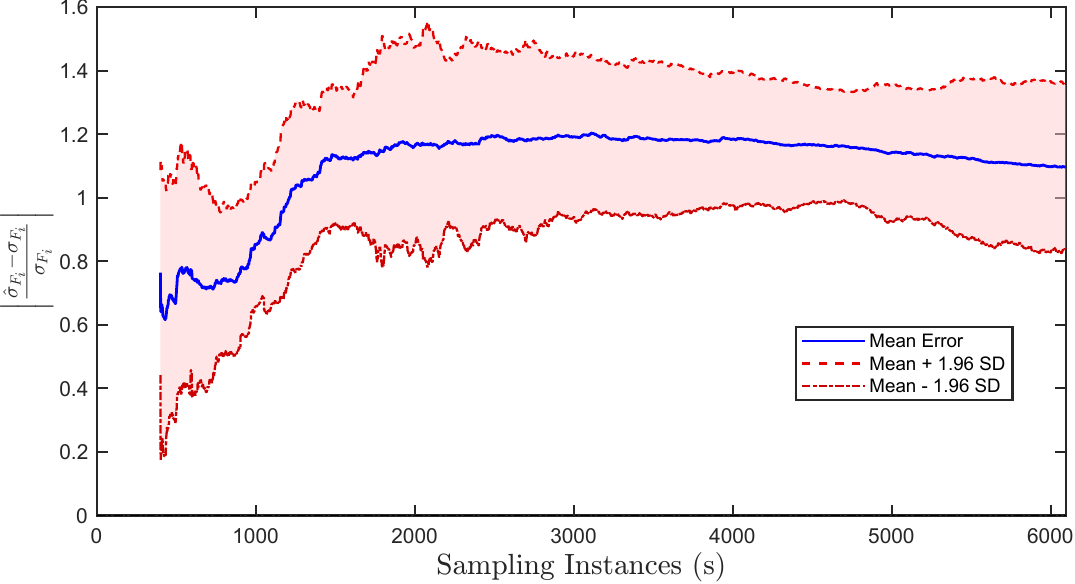}
        \subcaption[Short list entry]{The relative error in the input SD estimate shows an initial divergence, indicating biased estimate}
        \label{Figure_6c}
    \end{subfigure}
    \hfill
    \begin{subfigure}[t]{0.48\columnwidth}
        \centering
        \includegraphics[width=1.0\columnwidth]{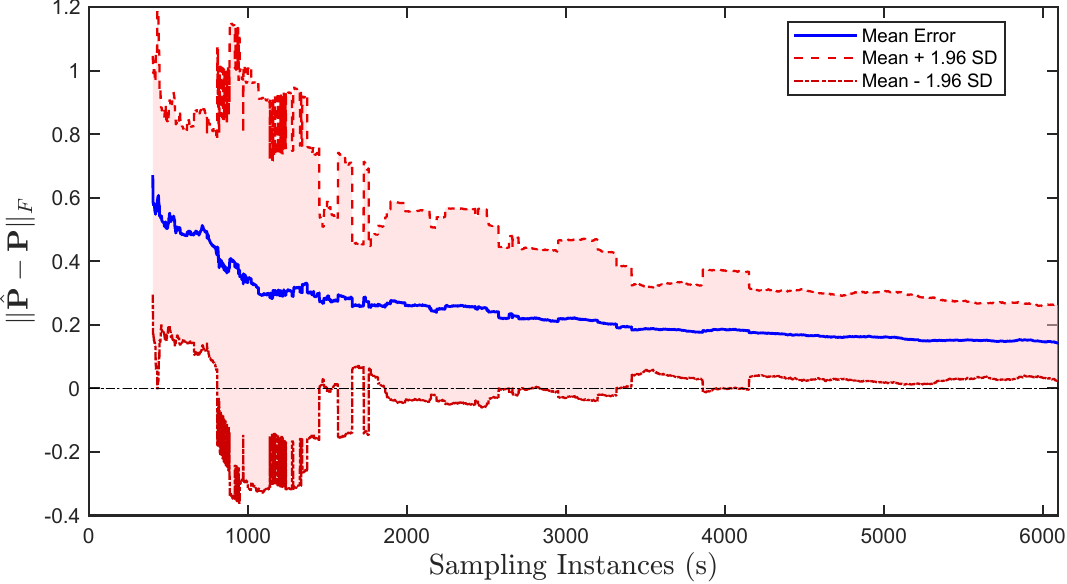}
        \subcaption[Short list entry]{Frobenius norm of the difference between the estimated and true poles of the time-varying system}
        \label{Figure_6d}
    \end{subfigure}
    
    \caption{Performance of RSMI-IPCA on the system, defined in \Cref{eq:4.4} under changing operating condition.}
    \label{Figure_6}
\end{figure}

\Cref{Figure_6a} shows that the process order estimate converges to true order within $3500$ samples. From \Cref{Figure_6b}, we observe that the relative error in the estimate of the output noise SD converges to zero as new measurements are received. However, in \Cref{Figure_6c}, we find an initial increment in the relative error in the input SD estimate, it starts to slowly reduce with the increasing number of samples, and require more samples to converge to the true value. The final estimate of the input noise SD is $\hat{\sigma}_{F_i} = 0.3314 \pm 0.0408$, which still has a bias and require more samples for RSMI-IPCA to converge to an estimate close to the corresponding true value, whereas the estimated output SD $\hat{\sigma}_{h_2} = 0.1314 \pm 0.0028$ is fairly accurate. 

Further, we track the poles of the system as it shifts from one steady-state operating point to another. The initial values of the true poles of the system are $(0.741, 0.714)$, which attain the updated values $(0.626, 0.919)$ from $1095$'th instant. Before the process order gets converged, the number of poles estimated from the model matrices might be different from the true number of poles, which is why we have padded either the vector containing true poles $\mathbf{P}$ or the vector $\mathbf{\hat{P}}$ containing the estimated poles with zeros and reported the Frobenius norm of the difference between the true poles and its estimates in \Cref{Figure_6d}. As we observe that the pole estimates gradually converges to the true values, with the final estimates being $0.615 \pm 0.194$ and $0.967 \pm 0.095$, which contain the respective true values, indicating the unbiased nature of these estimates. 

\subsubsection{Case study 3: Tracking changes in model structure} \label{subsubsec3:order}
In this study, we explore the ability of RSMI-IPCA to track changes in the process structure. We utilize the system defined in \Cref{eq:4.4}, with the identical values for the cross-sectional areas specified earlier, while keeping the valve coefficients fixed as $Cv_1 = 1.8,$ and $Cv_2 = 1.2$. Although, the input data generation is identical to the process followed in \Cref{subsubsec2:process} with a sample size of $N=6095$, initially we assume to have a level sensor attached only to tank 1. This gives the measurement of only the output variable $h_1(t)$, which we have generated using the \Cref{eq:4.4a} for the first $1495$ sampling instances. 

The changes in the process structure is introduced after $1495$ samples, by simultaneously generating the true measurements of output variable $h_2(t)$ using \Cref{eq:4.4b}. Therefore, for the remaining sampling instances, we have the measurements of $F_i, h_1,$ and $h_2$. In practice, this type of structural change might occur due to adding a new level sensor to the second tank to understand the process dynamics more accurately with more data. The true measurements are further corrupted using Gaussian white noise sequences with fixed standard deviations $\sigma_{F_i} = 0.0787,$ $\sigma_{h_1} = 0.0387,$ and $\sigma_{h_2} = 0.0592$. 

\begin{figure}[!ht]
    \centering

    \begin{subfigure}[t]{0.48\columnwidth}
        \centering
        \includegraphics[width=1.0\columnwidth]{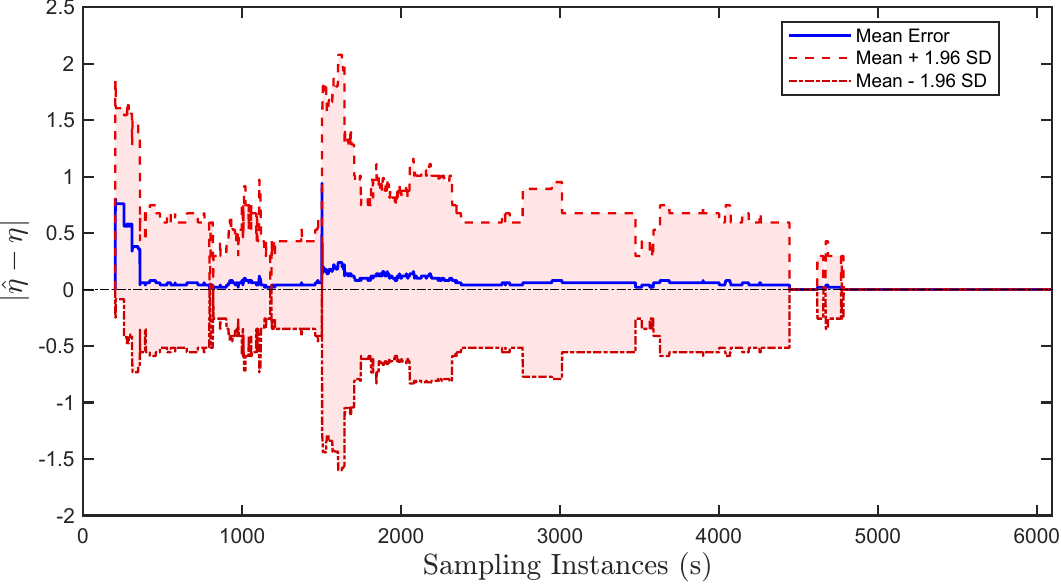}
        \subcaption[Short list entry]{Absolute difference between the estimated and true process orders}
        \label{Figure_7a}
    \end{subfigure}
    \hfill
    \begin{subfigure}[t]{0.48\columnwidth}
        \centering
        \includegraphics[width=1.0\columnwidth]{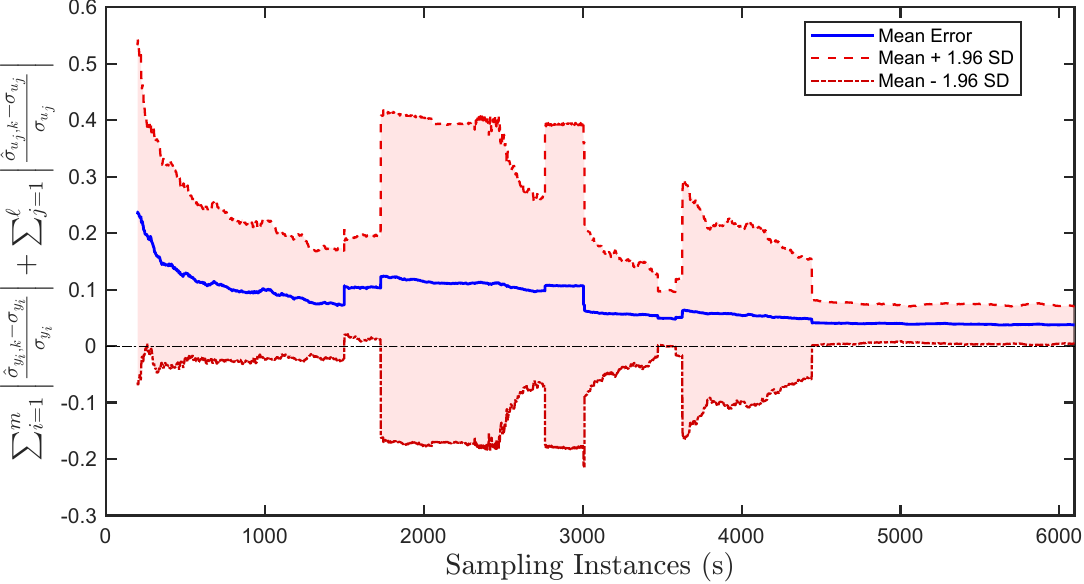}
        \subcaption[Short list entry]{Sum of the relative difference between the estimated and true SDs of the system variables}
        \label{Figure_7b}
    \end{subfigure}

    \vspace{0.8em}

    \begin{subfigure}[t]{0.48\columnwidth}
        \centering
        \includegraphics[width=1.0\columnwidth]{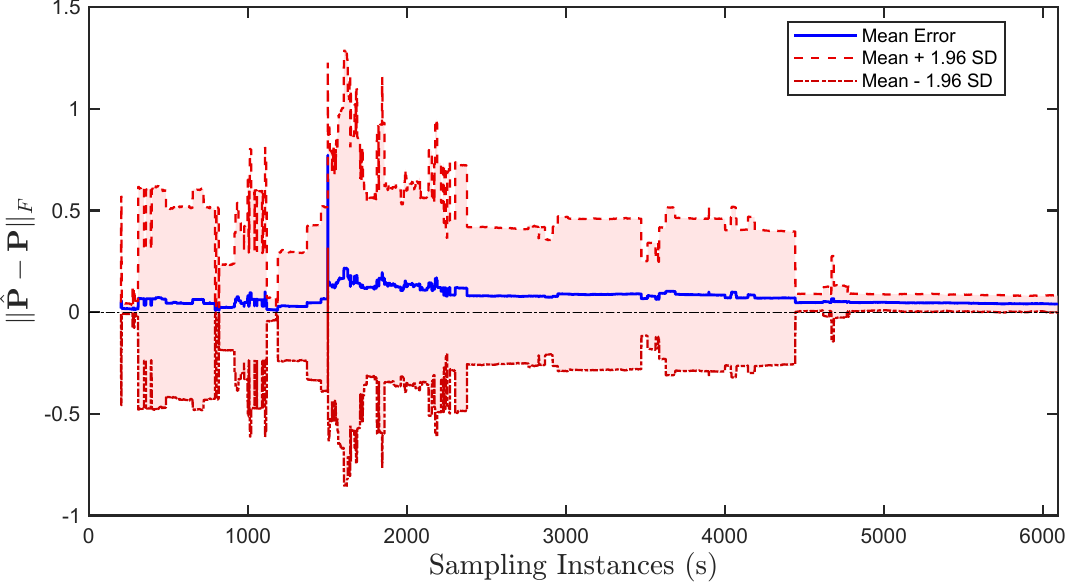}
        \subcaption[Short list entry]{Estimated poles are gradually converging to the true values as new measurements are received}
        \label{Figure_7c}
    \end{subfigure}
    
    \caption{Performance of RSMI-IPCA on the system, defined in \Cref{eq:4.4} in presence of a structural variation.}
    \label{Figure_7}
\end{figure}

We have carried out $50$ simulation trials with different noise realizations, where each trial consists of $N$ samples. The first $200$ samples (containing measurements of only $F_i,$ and $h_1$) are used to estimate the noise variances, process order, and model matrices using SMI-IPCA algorithm with a lag $f=6$, which are further used to initialize the RSMI-IPCA algorithm. Subsequently, we continue to apply RSMI-IPCA as and when new sample is received to update the noise variances, process order, and model matrices until the first $1495$ samples. After $1495$'th sample, as we start receiving the measurement of the other output variable $h_2$, we still continue to update the lagged sample covariance containing only $F_i,$ and $h_1$, along with storing the measurements of $h_2$ till $1500$'th sampling instant. This is to ensure that we have sufficient measurements of variable $h_2$ to construct the lagged data vector containing all the three variables. As we receive the $1501$'th sample, we extend the lagged sample covariance matrix to accommodate the variable $h_2$ by appropriately placing $1$ in the diagonals, whereas keeping the off-diagonal terms of the extended covariance matrix equal to zero. The initial estimate of the noise variance corresponding to the variable $h_2$ is chosen to be a small fraction of the $(2,2)$'th element of $(\mathbf{z}_f(k)\mathbf{z}_f(k)^\intercal)/1501,$ corresponding to the variable $h_2(k)$. The process order estimate is taken to be identical as before due to having no available knowledge of the effect of the changed model structure on the process order.  With these modified estimated, we apply RSMI-IPCA to update the estimates. The results of this experiment over $N=6095$ samples, are reported in \Cref{Figure_7}.

\Cref{Figure_7a} shows that the process order nearly converges to the true value $\eta = 1$ before $1501$'th instant, around which we observe a significant error due to introducing a structural change in the process. The system has shifted from a first-order process to a second order process. However, as we keep on receiving more samples of the three variables, the process order gradually converges to the true value $\eta=2$ within $4000$ samples. In \Cref{Figure_7b}, we observe the error in the noise SDs to converge to zero around $1500$ samples, as the confidence band contain zero within it. After this instant, we find a sudden increase in the relative error due to poor initial noise SD estimate of the newly introduced variable, which again converges to their respective true values as new measurements are kept on coming. The final estimates of the noise SDs are $\hat{\sigma}_{F_i} = 0.0788 \pm 0.0036,$ $\hat{\sigma}_{h_1} = 0.0390 \pm 0.0009,$ and $\hat{\sigma}_{h_2} = 0.0588 \pm 0.0010$. The $95\%$ confidence interval of the estimates contain their respective true values, indicating their unbiased nature. Study of the system poles also shows a gradual convergence before the introduction of the structural change around $1500$'th instant as shown in \Cref{Figure_7c}. Following this, even though the initial error in the estimated poles are significantly high, they are observed to come close to their respective true values as new measurements are kept on coming, yielding unbiased estimates as reported in \Cref{table:3}.

\begin{table}[!ht]
\centering
\begin{minipage}{\columnwidth}
\centering
\begin{threeparttable}
\caption{Estimates of poles of the system in \Cref{eq:4.4}.}
\label{table:3}
\renewcommand{\arraystretch}{1.25}
\begin{tabular}{cccc}
\toprule
True poles $(p)$ & $\mu\left( \hat{p} \right)$ & $1.96 \times \sigma\left( \Re(\hat{p}) \right)$ & $1.96\times \sigma\left( \Im(\hat{p}) \right)$ \\
\midrule
$0.741$ & $0.754 + 0.007i$ & $0.039$ & $0.030$ \\
$0.714$ & $0.698 - 0.008i$ & $0.048$ & $0.031$ \\
\bottomrule
\end{tabular}
\end{threeparttable}
\end{minipage}
\end{table}



\section{Conclusion and future work} \label{sec:conclusion}
In this work, we proposed a recursive method called RSMI-IPCA, for estimating the time-varying model parameters, model order of linear state space model of a MIMO process, together with the noise variances for the EIV case, in presence of both the process and measurement noises. This algorithm appropriately combines a recursive update procedure for the lagged data covariance matrix with concepts drawn from previously developed Recursive IPCA and Subspace-based Identification methods. Use of a formal hypothesis test facilitates the monitoring of the changes in system order, or model structure induced by process operational changes, which is computationally more tractable compared to the use of any heuristic approach. This is demonstrated by the simulation studies, which shows the successful tracking of the changes in the measurement noise variances due to sensor degradation. The successful adaptation in the model parameters is presented by tracking the poles and zeros of the dynamic system. A potential future work would be to extend the proposed approach to a more general scenario to adapt and identify the innovations form of the model, when the process noise is not of full rank. 

\appendix
\phantomsection
\refstepcounter{mainappendix}\label{app}
\section*{Appendix}
\addcontentsline{toc}{section}{Appendix}

\renewcommand{\thesubsection}{\Alph{subsection}}
\counterwithin{figure}{subsection} 
\counterwithin{table}{subsection}  
\counterwithin{equation}{subsection} 

We concisely put together two important aspects in this section, starting with the SMI-IPCA algorithm~\cite{Keerthan:2023} in presence of process noise in \Cref{app:algo}, which forms the basis of the core development of this paper. In \Cref{app:est}, we provide the detailed matrix algebra, required to estimate the state space model matrices from the estimated constraint model.

\subsection{SMI-IPCA algorithm in presence of process noise} \label{app:algo}
The SMI-IPCA algorithm is presented in \Cref{algo:2} which consists of four iterative loops. The outer level of iteration contains two loops. One for the hypothesis testing which is used to determine the order of the process, whereas the other outer loop runs till the convergence of $\mathbf{\hat{A}}_d$. The authors in \cite{Keerthan:2023} have specified that steps \ref{alg2:st19} to \ref{alg2:st24} are optional, yet recommended for improved estimates of system matrices $\mathbf{\hat{A}}$ and $\mathbf{\hat{C}}$. The second level of iteration consists of the loop, which runs till the convergence of $\mathbf{\hat{\Sigma}}_{\mathbf{e}f}$. This can be checked by the relative change in the sum of smallest $d_{guess}$ eigenvalues, which, if falls under the specified tolerance value, confirms the convergence of $\mathbf{\hat{\Sigma}}_{\mathbf{e}f}$. The last level of iteration, which is not explicitly shown, is the optimization of the defined nonlinear objective function in order to obtain converged estimates of $\mathbf{\hat{\Sigma}}_v$ and $\mathbf{\hat{\Sigma}}_w$, when provided with $\mathbf{Z}_f$ and $\mathbf{\hat{G}}_f$.

\begin{algorithm}[!ht]
    \caption{Simultaneous estimation of state space model matrices and noise covariances}
    
    \footnotesize
    \KwIn{Input and output data matrices $\mathbf{Y} \in \mathbb{R}^{N\times m}$ and $\mathbf{U} \in \mathbb{R}^{N\times \ell}$, respectively, and a sufficiently large lag value $f$}
    \KwOut{Estimates of system matrices $(\mathbf{\hat A}, \mathbf{\hat B}, \mathbf{\hat C}, \mathbf{\hat D})$ and measurement noise covariances $\mathbf{\hat \Sigma}_w, \mathbf{\hat \Sigma}_v$}

        \vspace{0.5em}
        \nl\label{alg2:st1}Construct the lagged data matrix $\mathbf{Z}_f$ using \Cref{eq:2.19,eq:2.20}. Initialize $d_{guess} = d_{\max}$, which is general $mf-1$\;

        \nl\label{alg2:st2}\While{$d_{guess} \geq d_{\min}$} {
        
            \nl\label{alg2:st3}Initialize $\mathbf{\hat{A}}, \mathbf{\hat{C}},$ and compute $\mathbf{\hat{G}}_f$ using $\mathbf{\hat{A}},\mathbf{\hat{C}}$ based on \Cref{eq:2.18}\;
            
            \nl\label{alg2:st4}Make an initial guess for $\mathbf{\hat{\Sigma}}_v$ and $\mathbf{\hat{\Sigma}}_w$\;
            
            \nl\label{alg2:st5}\While{$\mathbf{\hat{\Sigma}}_{\mathbf{e}f}$ not converged} {
            
                \nl\label{alg2:st6}Construct $\mathbf{\hat{\Sigma}}_{\mathbf{e}f}$ from $\mathbf{\hat{G}}_f, \mathbf{\hat{\Sigma}}_v,$ and $\mathbf{\hat{\Sigma}}_w$ using \Cref{eq:2.28}\;
                
                \nl\label{alg2:st7}Scaled the lagged data as $\mathbf{Z_S}_f = \mathbf{Z}_f\mathbf{\hat{\Sigma}}_{\mathbf{e}f}^{-1/2},$ and perform SVD on $\mathbf{Z_S}_f$ to obtain right singular matrix $\mathbf{\hat{V}}_{\mathbf{S}f}$\;
                
                \nl\label{alg2:st8}Estimate the constraint matrix as $\mathbf{\hat{A}}_d = \left(\mathbf{\hat{V}}_{\mathbf{S}f}\right)_{d_{guess}}^\intercal \times \left(\mathbf{\hat{\Sigma}}_{\mathbf{e}f}^{-1/2}\right)$, where $\left(\mathbf{\hat{V}}_{\mathbf{S}f}\right)_{d_{guess}}$ denotes the $d_{guess}$ columns of $\mathbf{\hat{V}}_{\mathbf{S}f}$ corresponding to the $d_{guess}$ smallest eigenvalues\;
                
                \nl\label{alg2:st9}Compute $\mathbf{\hat{\Gamma}}_f$ from $\mathbf{\hat{A}}_d$ using \Cref{eq:app.B1}\;
                
                \nl\label{alg2:st10}Obtain the estimates $\mathbf{\hat{A}}, \mathbf{\hat{C}}$ from $\mathbf{\hat{\Gamma}}_f$ using \Cref{eq:app.B3a,eq:app.B3b}\;
                
                \nl\label{alg2:st11}Compute $\mathbf{\hat{G}}_f$ using $\mathbf{\hat{A}},\mathbf{\hat{C}}$ from \Cref{eq:2.18}\;
                
                \nl\label{alg2:st12}Optimize the objective function defined in \Cref{eq:2.29} to get new estimates of $\mathbf{\hat{\Sigma}}_v$ and $\mathbf{\hat{\Sigma}}_w$\;
            }
            \nl\label{alg2:st13}Employ hypothesis test~\cite{Keerthan:2023} for the equality of smallest $d_{guess}$ eigenvalues of $\mathbf{Z_S}_{f}$\;
            
            \nl\label{alg2:st14}\uIf{null hypothesis is rejected} {
            
                \nl\label{alg2:st15}$d_{guess} \leftarrow d_{guess}-1$ \Comment*[r]{Reduce it gradually}
            }
            \nl\label{alg2:st16}\Else {
                \nl\label{alg2:st17}break\;
            }
        }
        \nl\label{alg2:st18}Set $\hat{d} = d_{guess}$ and compute the system order $\hat{\eta}$ using \Cref{eq:2.31}\;
        
        \nl\label{alg2:st19}Reconfigure $\mathbf{Z}_f$ with $f = \hat{\eta} +1$ and recompute $\mathbf{\hat{G}}_f$ \Comment*[r]{Optional steps}
        
        \nl\label{alg2:st20}\While{$\mathbf{\hat{A}}$ not converged} 
        {
            \nl\label{alg2:st21}Compute $\mathbf{\hat{\Sigma}}_{\mathbf{e}f}$ using $\mathbf{\hat{\Sigma}}_v, \mathbf{\hat{\Sigma}}_w$ and $(\mathbf{\hat{A}}, \mathbf{\hat{C}})$ based on \Cref{eq:2.18,eq:2.27b,eq:2.27c,eq:2.28}\;
            
            \nl\label{alg2:st22}Estimate $\mathbf{\hat{A}}_d$ from the eigenvectors of $\mathbf{Z_S}_f = \mathbf{Z}_f \mathbf{\hat{\Sigma}}_{\mathbf{e}f}^{-1/2}$\;
            
            \nl\label{alg2:st23}Compute $\mathbf{\hat{\Gamma}}_f$ from $\mathbf{\hat{A}}_d$ using \Cref{eq:app.B1}\;
            
            \nl\label{alg2:st24}Compute $\mathbf{\hat{A}}, \mathbf{\hat{C}}$ from $\mathbf{\hat{\Gamma}}_f$ using \Cref{eq:app.B3a,eq:app.B3b}\;
        }

        \nl\label{alg2:st25}Compute $\mathbf{\hat{H}}_{f1}$ from $\mathbf{\hat{\Gamma}}_f$ and $\mathbf{\hat{A}}_d$ using \Cref{eq:app.B6}\;
        
        \nl\label{alg2:st26}Estimate $\mathbf{\hat{B}}, \mathbf{\hat{D}}$ from $\mathbf{\hat{\Gamma}}_f, \mathbf{\hat{H}}_{f1}$ from \Cref{eq:app.B7a,eq:app.B7b}\;
        
    \label{algo:2}
\end{algorithm}

\subsection{Estimation of state space model matrices} \label{app:est}
Once an estimate of the dynamic constraint matrix $\mathbf{\hat{A}}_d$ is obtained, the subspace matrices $\mathbf{A},$ $\mathbf{B},$ $\mathbf{C},$ and $\mathbf{D}$ can be estimated, which starts with extracting $\mathbf{\hat{A}}_{d,y}$ and $\mathbf{\hat{A}}_{d,u}$, the sub-matrices of $\mathbf{\hat{A}}_d$ corresponding to the lagged outputs and inputs, respectively. 
\begin{align}
    &\mathbf{\hat{A}}_{d,y} = \mathbf{\hat{A}}_d(:,\ 1:mf) \nonumber \\ &\implies \mathbf{\hat{\Gamma}}_f^\perp = \mathbf{\hat{A}}_{d,y}^\intercal \label{eq:app.B1} \\
    &\mathbf{\hat{A}}_{d,u} = \mathbf{\hat{A}}_d(:,\ (mf+1):(m+\ell)f) \nonumber \\ &\implies -\mathbf{\hat{H}}_f^\intercal \mathbf{\hat{\Gamma}}_f^\perp = \mathbf{\hat{A}}_{d,u}^\intercal \label{eq:app.B2}
\end{align}

The extended observability matrix $\mathbf{\hat{\Gamma}}_f$ is computed as the null space of $\mathbf{\hat{\Gamma}}_f^\perp$. Consequently, the matrices $\mathbf{A},$ $\mathbf{C}$ can be estimated by exploiting the structure of $\mathbf{\hat{\Gamma}}_f$ defined in \Cref{eq:2.16}, using following two relations:
\begin{subequations}
    \label{eq:app.B3}
    \begin{align}
        &\mathbf{\hat{C}} = \mathbf{\hat{\Gamma}}_f(1:m,\ :) \label{eq:app.B3a} \\
        &\mathbf{\hat{\Gamma}}_f(1:m(f-1),\ :)\ \mathbf{\hat{A}} = \mathbf{\hat{\Gamma}}_f(m+1:mf,\ :) \label{eq:app.B3b}
    \end{align}
\end{subequations}

The structure of $\mathbf{H}_f$ as defined in \Cref{eq:2.17} is further exploited to estimate $\mathbf{B}$ and $\mathbf{D}$. Thereby, the first column of $\mathbf{H}_f$, denoted as $\mathbf{H}_{f1}$, is written as:
\begin{equation}
    \label{eq:app.B4}
    \mathbf{H}_{f1} = 
        \begin{bmatrix}
            \mathbf{D} \\ \mathbf{CB} \\ \vdots \\ \mathbf{CA}^{f-2}\mathbf{B}
        \end{bmatrix}
\end{equation}
The columns of matrices $\mathbf{\hat{A}}_{d,y}$ and $\mathbf{\hat{A}}_{d,u}$ are further partitioned and written as follows:
\begin{subequations}
    \label{eq:app.B5}
    \begin{align}
        &-\mathbf{\hat{A}}_{d,y} = 
            \begin{bmatrix}
                \mathbf{\Phi}_1 & \mathbf{\Phi}_2 & \ldots & \mathbf{\Phi}_f   
            \end{bmatrix}; \\
        &\mathbf{\hat{A}}_{d,u} = 
            \begin{bmatrix}
                \mathbf{\Psi}_1 & \mathbf{\Psi}_2 & \ldots & \mathbf{\Psi}_f   
            \end{bmatrix}
    \end{align}
\end{subequations}
where, $\mathbf{\Phi}_i: (mf-\eta)\times m$ and $\mathbf{\Psi}_i:(mf-\eta)\times \ell,$ for $i=1,\ldots,f$. In order to obtain $\mathbf{\hat{H}}_{f1}$, we exploit the structure of $\mathbf{H}_f$ and by using the \Cref{eq:app.B1,eq:app.B2,eq:app.B4,eq:app.B5}, we obtain:
\begin{align}
    &-\mathbf{\hat{A}}_{d,y}\mathbf{\hat{H}}_f = \mathbf{\hat{A}}_{d,u} \nonumber \\
    &\implies\begin{bmatrix}
        \mathbf{\Phi}_1 & \mathbf{\Phi}_2 & \ldots & \mathbf{\Phi}_f   
    \end{bmatrix}
    \begin{bmatrix}
        \mathbf{\hat{H}}_{f1} & \ldots & \mathbf{\hat{H}}_{ff} 
    \end{bmatrix} = \mathbf{\hat{A}}_{d,u} \nonumber \\
    &\implies \begin{bmatrix}
        \mathbf{\Phi}_1 & \mathbf{\Phi}_2 & \ldots & \mathbf{\Phi}_{f-1} & \mathbf{\Phi}_f \\
        \mathbf{\Phi}_2 & \mathbf{\Phi}_3 & \ldots & \mathbf{\Phi}_f & \mathbf{0} \\
        \vdots & \vdots & \ddots & \vdots & \vdots \\
        \mathbf{\Phi}_f & \mathbf{0} & \ldots & \mathbf{0} & \mathbf{0} \\
    \end{bmatrix} \mathbf{\hat{H}}_{f1} = 
    \begin{bmatrix}
        \mathbf{\Psi}_1 \\ \mathbf{\Psi}_2 \\ \vdots \\ \mathbf{\Psi}_f   
    \end{bmatrix} \label{eq:app.B6}
\end{align}
Finally, the estimates $\mathbf{\hat{B}},$ $\mathbf{\hat{D}}$ are obtained using following two relations derived from the structure of $\mathbf{H}_{f1}$:
\begin{subequations}
    \label{eq:app.B7}
    \begin{align}
        &\mathbf{\hat{D}} = \mathbf{\hat{H}}_{f1}(1:m,\ :) \label{eq:app.B7a} \\
        &\mathbf{\hat{\Gamma}}_f(1:m(f-1),\ :)\ \mathbf{\hat{B}} = \mathbf{\hat{H}}_{f1}(m+1:mf,\ :) \label{eq:app.B7b}
    \end{align}
\end{subequations}

It may be noted that \Cref{eq:app.B3b,eq:app.B6,eq:app.B7} are over-determined systems of equations which are solved using least squares method. Therefore, the estimated system matrices from these equations are expected to be a transformed versions of the original system matrices, implying the inability to directly compare these matrices. However, the poles and zeros of the estimated and true system matrices, which are invariant properties of the system, are compared to assess the estimation accuracy. 

\providecommand{\latin}[1]{#1}
\makeatletter
\providecommand{\doi}
  {\begingroup\let\do\@makeother\dospecials
  \catcode`\{=1 \catcode`\}=2 \doi@aux}
\providecommand{\doi@aux}[1]{\endgroup\texttt{#1}}
\makeatother
\providecommand*\mcitethebibliography{\thebibliography}
\csname @ifundefined\endcsname{endmcitethebibliography}  {\let\endmcitethebibliography\endthebibliography}{}


\end{document}